\documentclass{aa}  
\usepackage{graphicx}
\usepackage{txfonts}
\usepackage{booktabs} 
\usepackage{pdflscape}
\usepackage{xcolor}
\usepackage{lscape}
\usepackage{float}
\usepackage[hidelinks]{hyperref}
   \hypersetup{
      colorlinks = true,
      citecolor ={cyan}
   }
\usepackage{longtable}
\usepackage{caption} 
\usepackage{subcaption}

\newcommand{\Msun}
{\,\ensuremath{\mathrm{M}_\odot}\xspace}

\newcommand{\appropto}{\mathrel{\vcenter{
  \offinterlineskip\halign{\hfil$##$\cr
    \propto\cr\noalign{\kern2pt}\sim\cr\noalign{\kern-2pt}}}}}

\begin{document}

   \title{A new sample of massive B-type contact binary candidates from the OGLE survey of the Magellanic Clouds} 


   \titlerunning{OGLE massive contact-binaries}

   \author{A. Menon\inst{1,2,3}, M. Pawlak\inst{4}, D. Lennon \inst{1}, K. Sen\inst{5,6}, N. Langer\inst{7,8}}
   \institute{Instituto de Astrofísica de Canarias, Avenida Vía Láctea s/n, 38205 La Laguna, Santa Cruz de Tenerife, Spain
    \and Universidad de La Laguna, Departamento de Astrofísica, Avenida Astrofísico Francisco Sánchez s/n, 38206 La Laguna, Santa Cruz de Tenerife, Spain
    \and  Department of Astronomy, 538 West 120th Street, Pupin Hall, Columbia University, New York City, NY 10027, U.S.A
    \and Lund Observatory, Division of Astrophysics, Department of Physics, Lund University, Box 43, SE-221 00, Lund, Sweden
    \and Institute of Astronomy, Faculty of Physics, Astronomy and Informatics, Nicolaus Copernicus University, Grudziadzka 5, 87-100 Torun, Poland
    \and Steward Observatory, Department of Astronomy, University of Arizona, 933 N. Cherry Ave, Tucson, AZ, 85721, USA
    \and Argelander Institut für Astronomie, Auf dem Hügel 71, DE-53121, Bonn, Germany
    \and Max-Planck-Institut für Radioastronomie, Auf dem Hügel 69, DE-53121, Bonn, Germany
    \\
    \email{aam2371@columbia.edu}}
   \date{}

\abstract
 {Massive contact binaries (CBs) are crucial objects for understanding close binary evolution and stellar mergers. Their study has been hampered by a scarcity of observed systems, particularly of B-type systems which are expected to dominate this class.}
   {We bridge this observational gap by mining a large sample of massive CB candidates from the OGLE-IV database, potentially increasing their current numbers in the Magellanic Clouds by an order of magnitude.}
   {Using main-sequence colour-magnitude limits, an observationally informed period-luminosity-colour relation for CBs, and a high morph parameter cut ($c\geq0.7$), we empirically identified a subsample of 68 O and B-type binaries with periods $P<3$\,days, that exhibit smooth, sinusoidal light curves with nearly equal eclipse depths. To mine our bona fide sample of CB candidates among these, we used theoretical colour-magnitude and orbital period distributions based on a vast grid of \texttt{MESA}  binary models. We also computed  synthetic light curves using \texttt{PHOEBE} corresponding to the contact and near-contact phases of a \texttt{MESA} model.} 
{ Our bona fide candidate CB sample consists of 37 systems (9 in the SMC and 28 in the LMC), which fulfil the theoretical predictions for massive CBs. The bona fide sample, which predominantly consists of B-type binaries with periods of $P \approx 0.6- 1$\,day, closely agrees with our predicted population count. As our binary models predict mass equalization followed by temperature equalization during nuclear-timescale contact, a substantial fraction of these bona fide CB candidates may have mass ratios of $q \approx 1$.}
   {Our work significantly expands the observational sample of B-type candidate massive CBs. Furthermore, our synthetic light curves show a degeneracy between contact and near-contact binary light curves, indicating the possibility of misidentifications between these configurations when characterized based on light curves alone. Spectroscopic follow-up is necessary to test our predictions, particularly for the mass ratios of the massive CB candidates.}
 
   \keywords{Stars: massive -- Stars: evolution -- Stars: early-type -- binaries: close -- binaries: eclipsing}

   \maketitle

\section{Introduction}
\label{introduction}

Contact binaries (CBs) are a critical class of massive stars that allow us to examine how the closest binary stars evolve and what their end fates may be. In CBs, both stars overflow their Roche lobe volumes simultaneously and share a common envelope over nuclear timescales, rendering the contact phase as an observationally accessible phenomenon. At relatively high metallicities ($Z\geq$\,Z$_\text{SMC}$), they are expected to be the progenitors of stellar mergers of main-sequence (MS) binaries \citep{Menon2021}, while at lower metallicities, CBs may  undergo chemically homogeneous evolution and become progenitors of black-hole  binaries \citep{Demink2009,Marchant2016}. As members of triple systems at low metallicities, the CB channel can further contribute to the population of black-hole binaries \citep{dorozsmai2024,vigna-gomez2025}.

Intermediate luminosity transients such as luminous red novae (LRNe) are expected to be powered by stellar mergers, the most famous of which resulted from the merger of V1309 Sco, a known low-mass CB \citep{tylenda2011, tylenda2016}. More luminous LRNe may hence arise from the mergers of massive CBs. Thus, determining the number of CBs and their properties in an observable population allows us to constrain the rate of MS stellar mergers and LRNe, and provides valuable tests for the formation pathways of compact-object binaries.

\citet{Menon2021} (hereafter Paper I) built  synthetic populations for the Large and Small Magellanic Clouds (LMC; SMC) from large grids of detailed binary models. They determined that massive CBs in the Clouds will be MS binaries with mass ratios of $q\approx1$ and predominantly consisting of B-type systems with periods $\lessapprox1$\,day, due to their long contact durations and higher weight in the initial mass function (IMF). This theoretical probabilistic distribution of orbital parameters  (P--q)  was found to be  similar within the metallicity range of the Magellanic Clouds.  

 Paper I also compiled the first comprehensive list of spectroscopically confirmed massive CBs and near-CBs from literature, i.e, those with radial velocity (RV) curves and light curves, which comprised of 26 systems in total, from the SMC, LMC and Milky Way (MW). The majority of the spectroscopic CBs were O-type systems with $T_\textrm{eff}\gtrapprox30$\,kK with periods $>1$\,d; B-type CBs were under reported. The majority of these belonged to the MW; only two confirmed CBs were identified in the Magellanic Clouds. 

\citet{abdul-masih2022} pioneered to determine orbital period-change rates from archival records for six O+O systems flagged in Paper I and thereby determine their mass-transfer rates; however, these estimates were limited by large uncertainties. More recently, \citet{vrancken2024} performed a similar analysis for five B+B systems, determining period-change rates with improved precision, and found values broadly consistent with the predictions of Paper I. A similar study had also been carried out by \citet{qian2007} for two massive CBs. These works concluded, within their uncertainties,  that a subset of massive CB candidates are undergoing nuclear-timescale mass transfer.

The field of massive CBs, thus, suffers from small-number statistics. To rigorously test theoretical predictions, a substantially larger, homogeneously analysed sample of massive CBs is required. For this, we use the largest photometric survey database of eclipsing binaries-- the Optical Gravitational Lensing Experiment (OGLE)-IV survey  of the Magellanic Clouds  \citep{Udalski2015,Pawlak2016}. This survey comprises 40,204 eclipsing and ellipsoidal binaries in the LMC and 8,401 in the SMC, and is largely complete for O and B-type MS  binaries.

In this paper, we first empirically select a sample of O- and B-type MS binaries from the OGLE database that have smooth, sinusoidal light curves with nearly equal depths, as expected for CBs with equal-temperature components. We further refine this sample based on our theoretical constraints and mine a ``bona fide'' subset that we expect contains the systems likeliest to be CBs-- 37 in total across the LMC and SMC-- thus increasing their known numbers in the Magellanic Clouds by an order of magnitude and making this the largest homogeneous sample of CB candidates reported to date. The majority of this sample consists of B-type systems with periods of 0.6--1\,days. In addition, we investigate the light curves of semi-detached and detached binaries nearing contact, compare them to those of CBs, and highlight potential observational misidentifications between these binaries.

This paper is divided as follows: Section~\ref{sample} describes the selection of massive MS binaries from the OGLE database and our empirical subsample of interest, Section~\ref{models} presents an exemplary binary evolution model, the construction of synthetic light curves and populations,  Section~\ref{comparison}  compares the empirical subsample and theoretical predictions to determine our bona fide CB candidate sample, Section~\ref{discussion} discusses the implications and caveats of our methodology and results, and places them in context with previous studies, and finally, in Section~\ref{conclusions}, we enlist the conclusions of our work.

\section{Observational sample from OGLE}
\label{sample}

 \begin{figure}[h]
   \centering
  \includegraphics[width=1.02\linewidth]{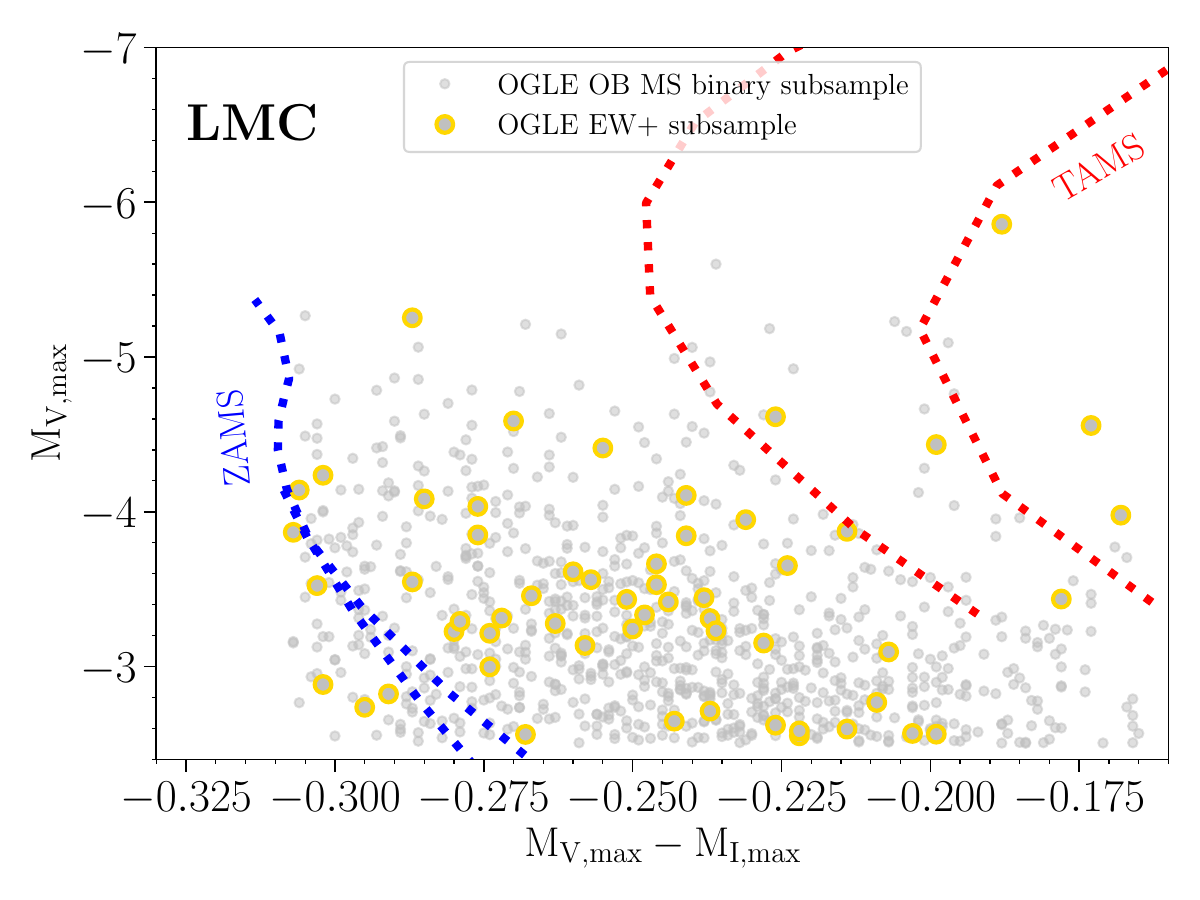}
   \caption{OB binaries with periods $P>3$\,days across all morph parameter values, identified as MS systems from the OGLE sample (light grey dots) on the absolute colour ($M_\textrm{V,max}-M_\textrm{I,max}$) -- magnitude ($M_\textrm{V,max}$) space (corresponding to the maximum brightness of the binary). Overlaid are the empirically identified OGLE EW+ subsample from the LMC (yellow circles, as described in Section~\ref{CB_sample}). Curves indicate the limits of Zero Age and Terminal Age Main Sequence (ZAMS and TAMS) curves computed from non-rotating and rotating ($v_\textrm{i}=330$\,km/s) models of  single stars with masses of $7-50$\Msun.}  
              \label{CMD_all}%
    \end{figure}

 \begin{figure}
    \centering
    \includegraphics[width=1.02\linewidth]{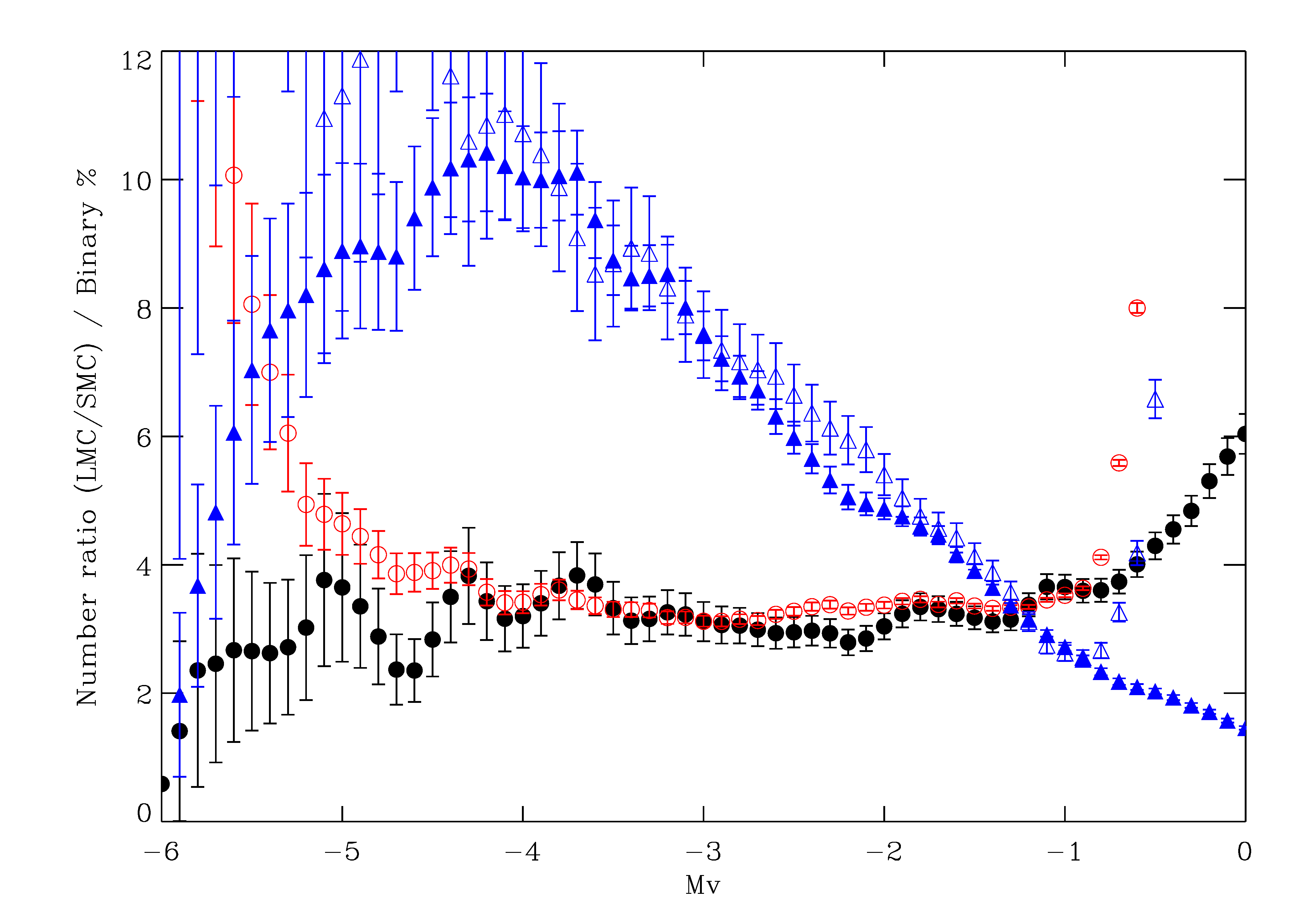}
    \caption{Characterization of the OGLE binary sample: Black filled circles represent the ratio of LMC to SMC OGLE binaries, red circles are the same ratio but for $Gaia$ MS stars, and blue triangles are the percentage binary fractions (OGLE/$Gaia$) for the LMC (filled triangles) and SMC (open triangles). The x-axis represents the absolute V-band magnitude $M_\textrm{V}$ (or $M_\textrm{V,max}$ in the case of OGLE binaries). Error bars represent Poisson uncertainties $\sqrt{N}$, where $N$ is the sample size. Absolute magnitudes were obtained by adopting distance moduli of 18.48 and 18.96 for the LMC and SMC respectively, with mean extinction corrections of 0.34 and 0.25 magnitudes.}
    \label{fig:ratios}
\end{figure}
 
 \subsection{The OGLE massive main-sequence binary sample}
 \label{MS_sample}
The OGLE survey has been monitoring the Magellanic Clouds for over two decades. The fourth and current phase of the project began in 2010 and covers an area of about 650 square degrees, including the LMC, the SMC, and the Bridge area between the two galaxies. The OGLE-IV observations are taken in Johnson's $V$ and Cousin's $I$ filters, with the majority (90\%) of the observations taken in $I$ band. The OGLE-IV saturation limit in the $I$ band is about 13~mag in the Magellanic Clouds. The survey has high completeness up to about 19~mag, however, some observations reach as faint as 21.5~mag. The technical details of the OGLE-IV survey can be found in \citet{Udalski2015}. 

The OGLE catalogue reports the apparent V- and I-band magnitudes corresponding to the maximum brightness of the binary, at the peak amplitude of its light curve. We define $M_{\text{V,max}}$ and $M_{I,\text{max}}$ as the absolute V- and I-band magnitudes of the binary at its light curve maximum, i.e., when the system is at peak brightness and both stars are most visible). First, we de-redden the individual OGLE sources using the optical reddening maps by \citet{Skowron2021}. Comparing the values of $E(V-I)$ implied by \citet{Skowron2021} for O-types stars in the LMC and SMC \citep[selected from][and other literature sources]{bonanos2009,bonanos2010} we find small mean underestimates of 0.016~mag in the LMC and 0.0125~mag in the SMC compared to the spectroscopic reddening values. Therefore, assuming these offsets are also appropriate for massive binaries, we correct $E(V-I)$ by these values, before de-reddening the sample. We adopt $R_\mathrm{V,LMC} = 3.4$ for the LMC and $R_\mathrm{V,SMC} = 2.74$ for the SMC \citep{gordon2003}.

Next, we correct the de-reddened magnitudes for the distance modulus ($DM$) in order to obtain the absolute magnitudes. For this purpose, we adopt the $DM_\mathrm{LMC}= 18.476$~mag for the LMC \citep{Pietrzynki2019} and $DM_\mathrm{SMC}= 18.997$~mag for the SMC \citep{Graczyk2020}. Stars in the Clouds are generally too far away to have good parallax measurements in Gaia, so any object that has a positive parallax and  parallax/parallax-error > 5 must be a Galactic foreground object and not an LMC or SMC member. We use Gaia DR3 parallax data \citep{Gaia2021} in order to remove these foreground objects from our sample.

We filter the massive MS binaries from this de-reddened sample using Zero Age Main Sequence (ZAMS) and Terminal Age Main Sequence (TAMS) CMD limits. These limits are determined using absolute V-band magnitudes and absolute V-I colours for LMC single-star models with masses ranging from 7 to 50 solar masses ($M_\odot$) and two initial rotational velocities: $v_i = 0$ km/s and $v_i = 330$ km/s, computed using \texttt{MESA} v. r15140 \citep{paxton2011,paxton2013,paxton2015,paxton2018}, which provides only bolometric luminosities. To convert these to absolute V and I-band magnitudes, we employ the bolometric corrections and colours from the TLUSTY O- and B-star model atmosphere grids \citep{lanz2003,lanz2007}.

Fig.~\ref{CMD_all} illustrates the resulting absolute colour-magnitude diagram (CMD). The theoretical ZAMS and TAMS curves of the MS tracks on the CMD indicate that massive MS binary systems containing stars with initial masses $7\Msun$ have  absolute V-band magnitudes of $M_\textrm{V}\gtrapprox-2.5$, absolute V-I colours of $-0.307\lessapprox M_\textrm{V} - M_\textrm{I}\lessapprox-0.165$.  We expect the OGLE binaries within the above limits to be massive MS binaries, corresponding roughly to spectral types B2 V and earlier according to the classification scheme of \citet{Pecaut2013}. Some points fall outside these limits; however, cross-matching the OGLE sample with known OB systems indicates that these are likely the result of incorrect extinction corrections derived from the \citet{Skowron2021} maps that are of low spatial resolution. In total, we obtain 2811 and 861 early MS binaries in the LMC and SMC respectively with orbital periods $P<30$\,days.

We also examine  the binary number statistics and 
completion of our sample selection for CBs.  For this, we use the $Gaia$ catalog to estimate the total number of MS stars in each galaxy that lie within the OGLE binary survey spatial and colour-magnitude footprints \citep[for a discussion of the completeness of $Gaia$ see][]{schoot2021}. 
Reassuringly, both the LMC-to-SMC ratios of OGLE binaries and $Gaia$ MS are nearly identical over an absolute magnitude range corresponding to spectral types O9/B0\,V ($M_\textrm{V}\sim-5$) to approximately B7\,V ($M_\textrm{V}\sim-1$), as seen in Fig.~ \ref{fig:ratios}. For the brightest magnitudes the OGLE survey suffers from incompleteness close to the saturation limits (around $M_\textrm{V,max}\sim-6.0$ in the LMC and $M_\textrm{V,max}\sim-6.4$ in the SMC) and stochastic effects (small numbers of stars). 
In addition, incompleteness due to faint limits impacts both OGLE and $Gaia$ at different absolute magnitudes. Assuming detection biases are roughly the same in OGLE for both LMC and SMC, Fig.\ref{fig:ratios} implies that binary fractions of late O- and B-type stars in both galaxies are roughly the same, indicating a metallicity independence in their distribution. This is also consistent with the findings of Paper I, who found a similar result for the population of CBs in both Magellanic Clouds.
While the ratios are similar, the trend of OGLE binary fraction relative to $Gaia$ MS stars exhibits a drop from around 10\% to a few percent at fainter magnitudes. We expect that this decrease reflects an observational bias in that lower signal-to-noise light curves make binary detection more difficult, and potentially an intrinsic decrease of binary fraction with decreasing mass \citep{banyard2022}. 

The average recall of the classifier used in \citet{Pawlak2016} is about 80\%, and is likely to be slightly higher for the early-type contact binaries, as they are bright and therefore have lower scatter than most of the sample. The main limiting factor for the completeness is the inclination of the system. Roughly half of the population of CBs may not show eclipses due to low inclination of the orbit. Some of these systems may be detected as ellipsoidal binaries (ELLs), though. A second limiting factor is the OGLE saturation limit, which excludes the brightest O-type stars (apparent magnitude $\lessapprox13.2$).
\subsection{Empirical OGLE EW+ subsample}
\label{CB_sample}

     \begin{figure*}
   \centering 
   \begin{subfigure}[b]{\textwidth}
   \centering
   \includegraphics[clip=true, trim={0 0.5cm 0 0},angle=270,width=0.34\linewidth]{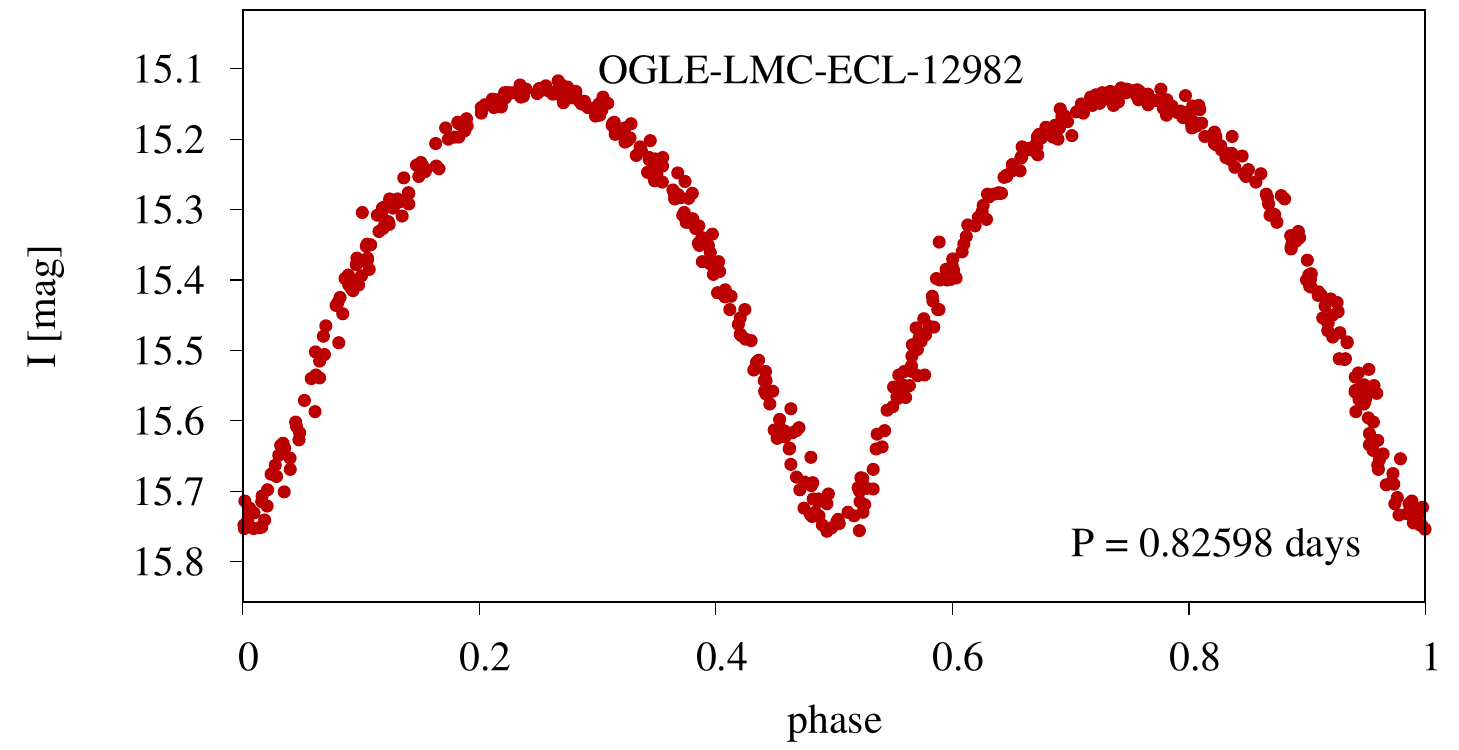}            \hfill
    \includegraphics[clip=true, trim={0 2cm 0 0}, angle=270,width=0.32\linewidth]{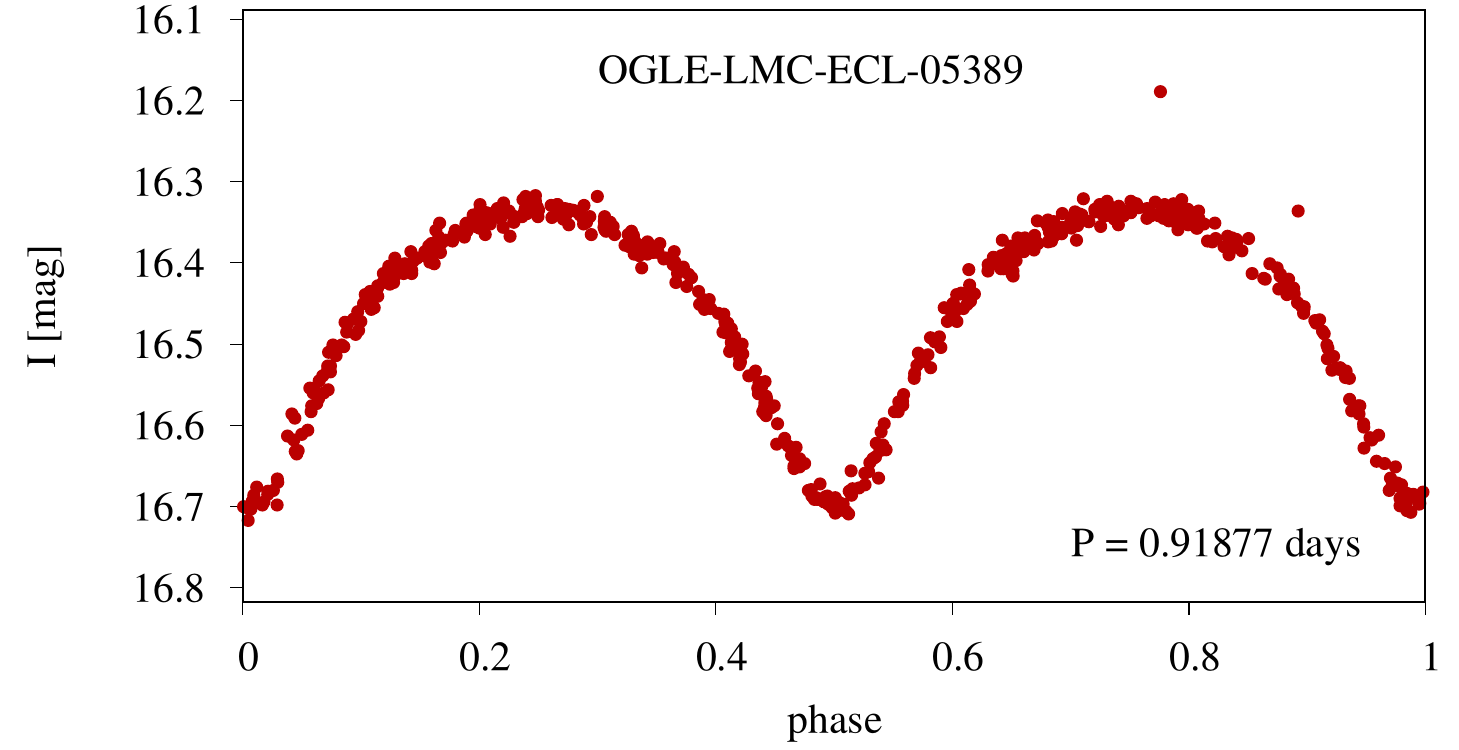}
    \hfill
    \includegraphics[clip=true, trim={0 2cm 0 0}, angle=270,width=0.32\linewidth]{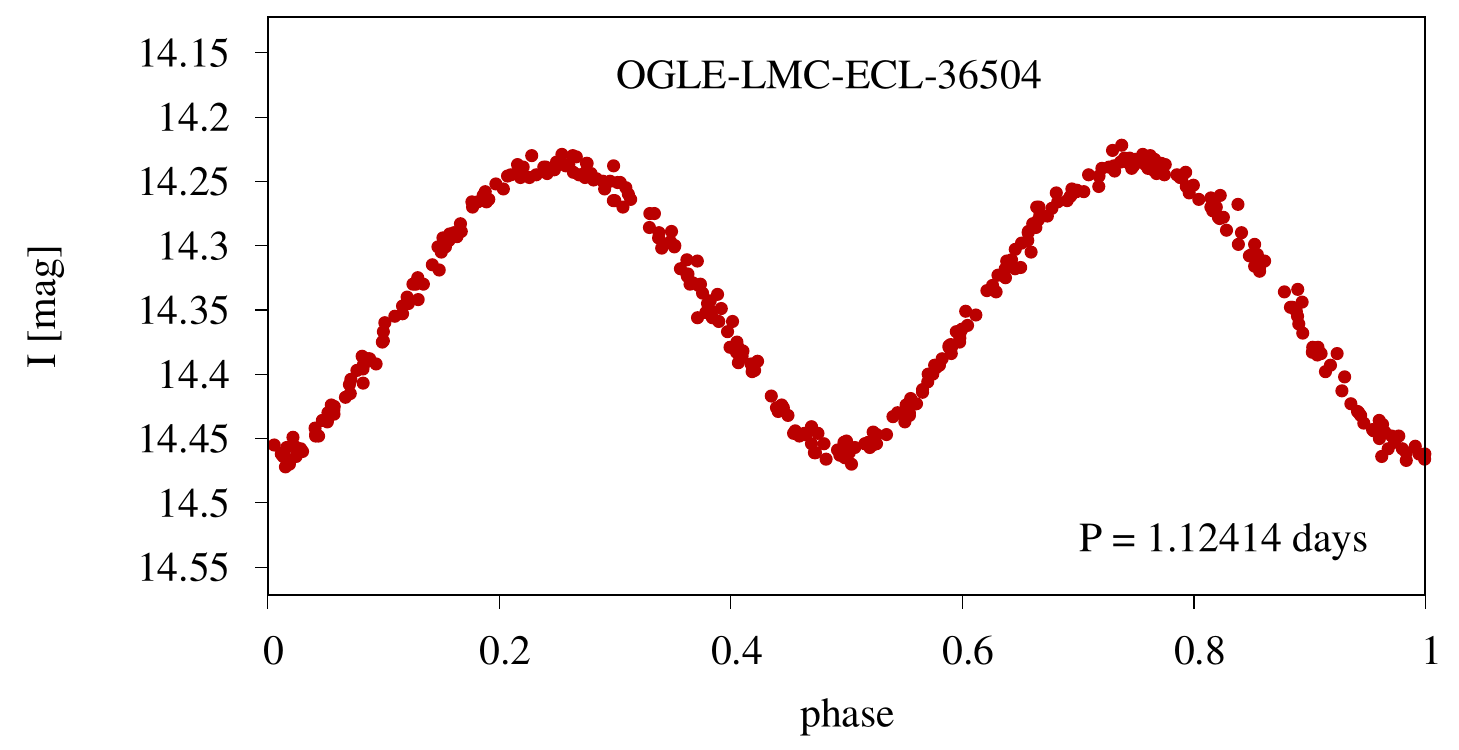}
    \hfill
     \end{subfigure}
   \begin{subfigure}[b]{\textwidth}
   \centering    
    \includegraphics[clip=true, trim={0 0.5cm 0 0},angle=270,width=0.34\linewidth]{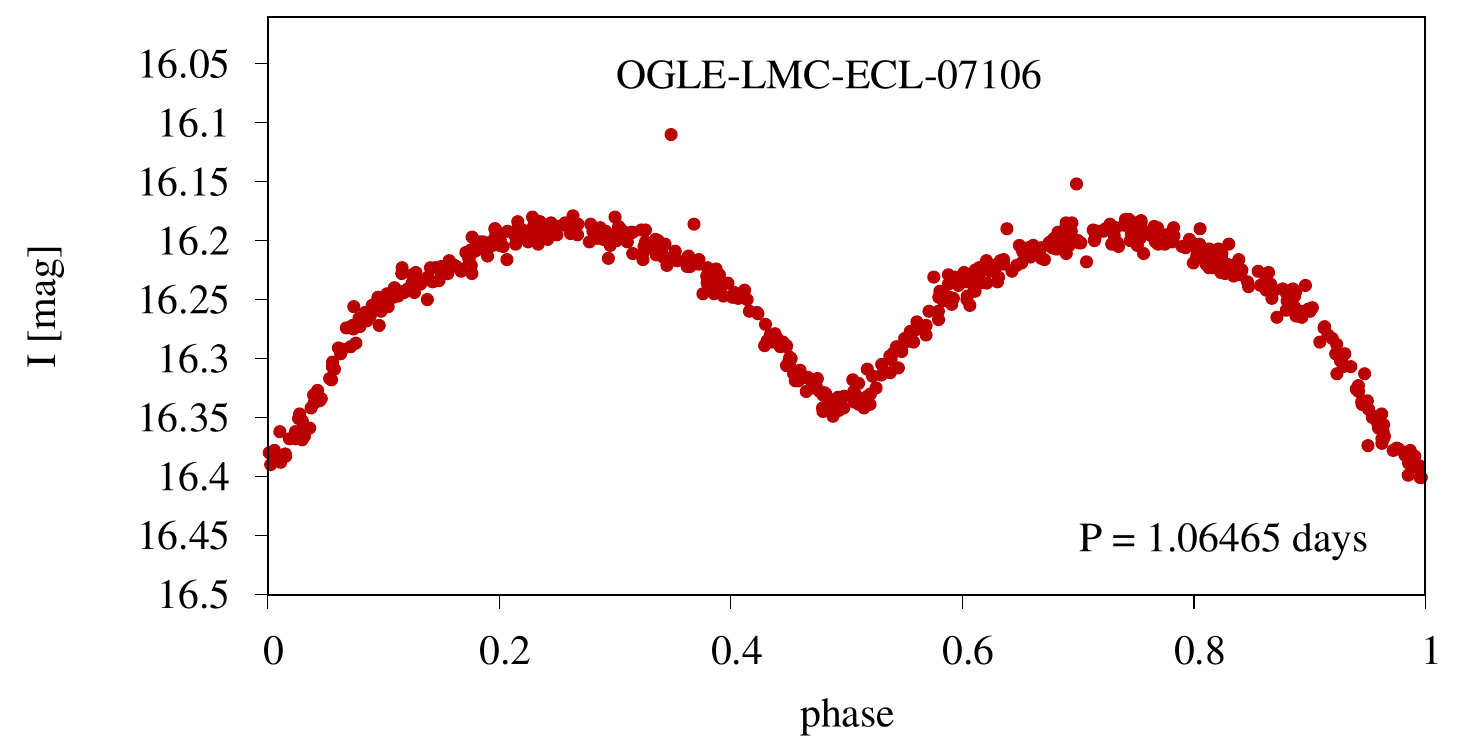}
    \includegraphics[clip=true, trim={0 2cm 0 0}, angle=270,width=0.32\linewidth]{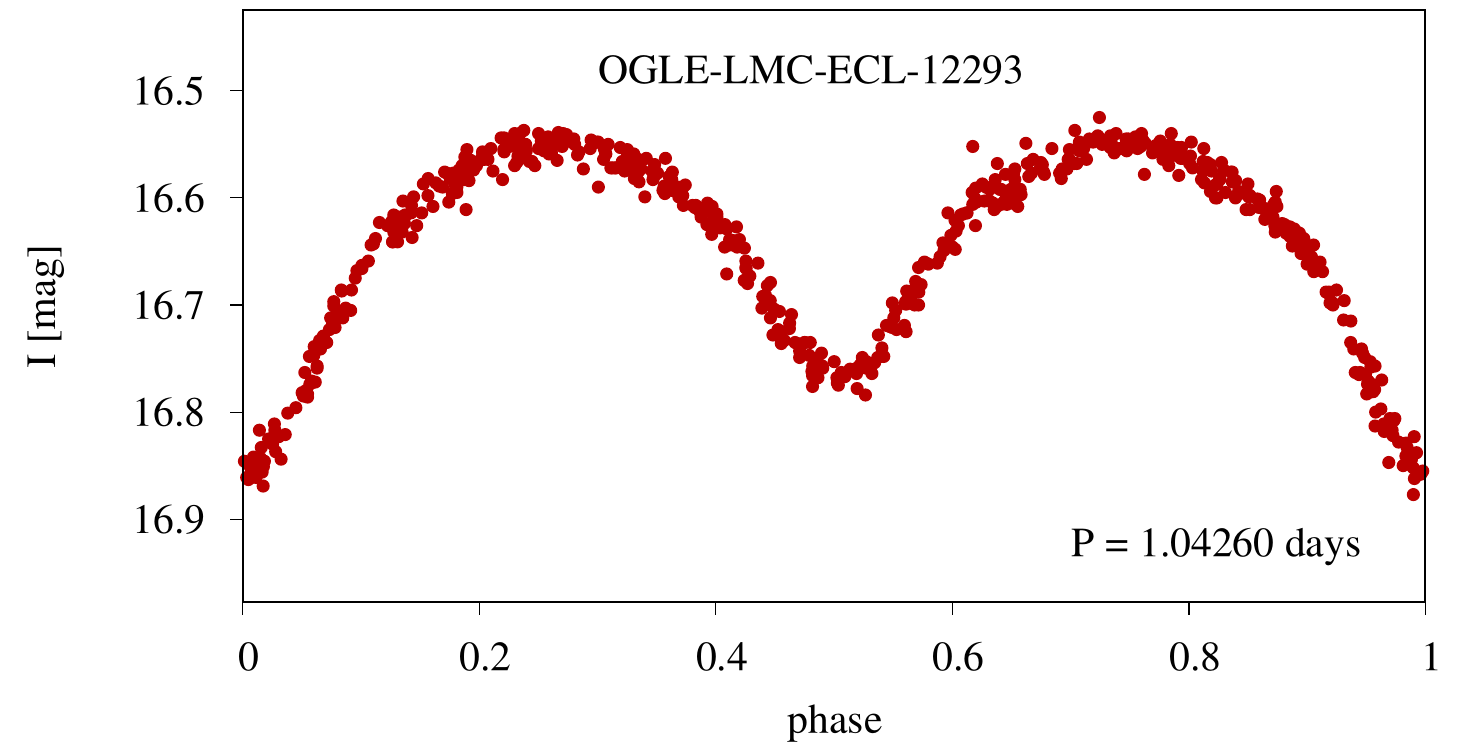}
    \includegraphics[clip=true, trim={0 2cm 0 0}, angle=270,width=0.32\linewidth]{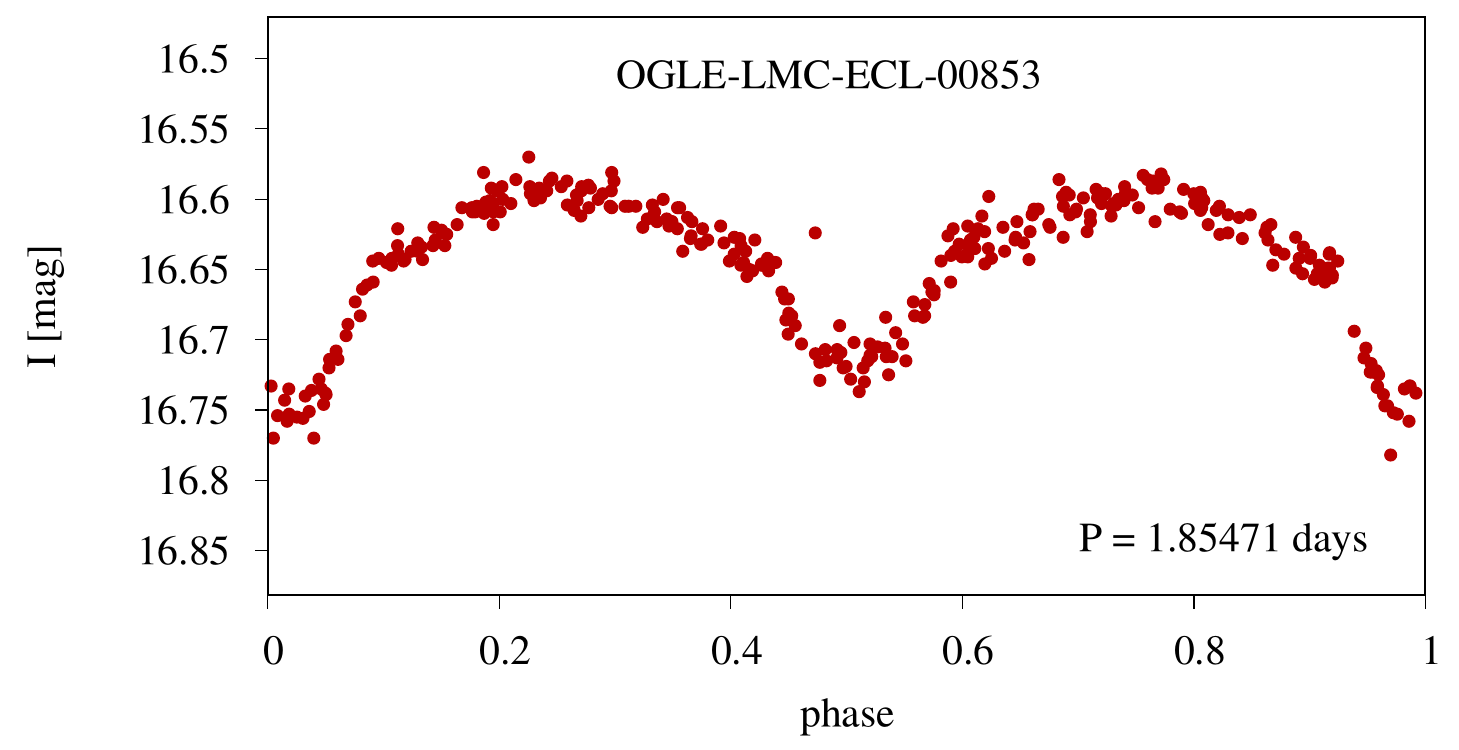}\\
    \end{subfigure}
   \begin{subfigure}[b]{\textwidth}
   \centering        
   \includegraphics[clip=true, trim={0 0.5cm 0 0}, angle=270,width=0.34\linewidth]{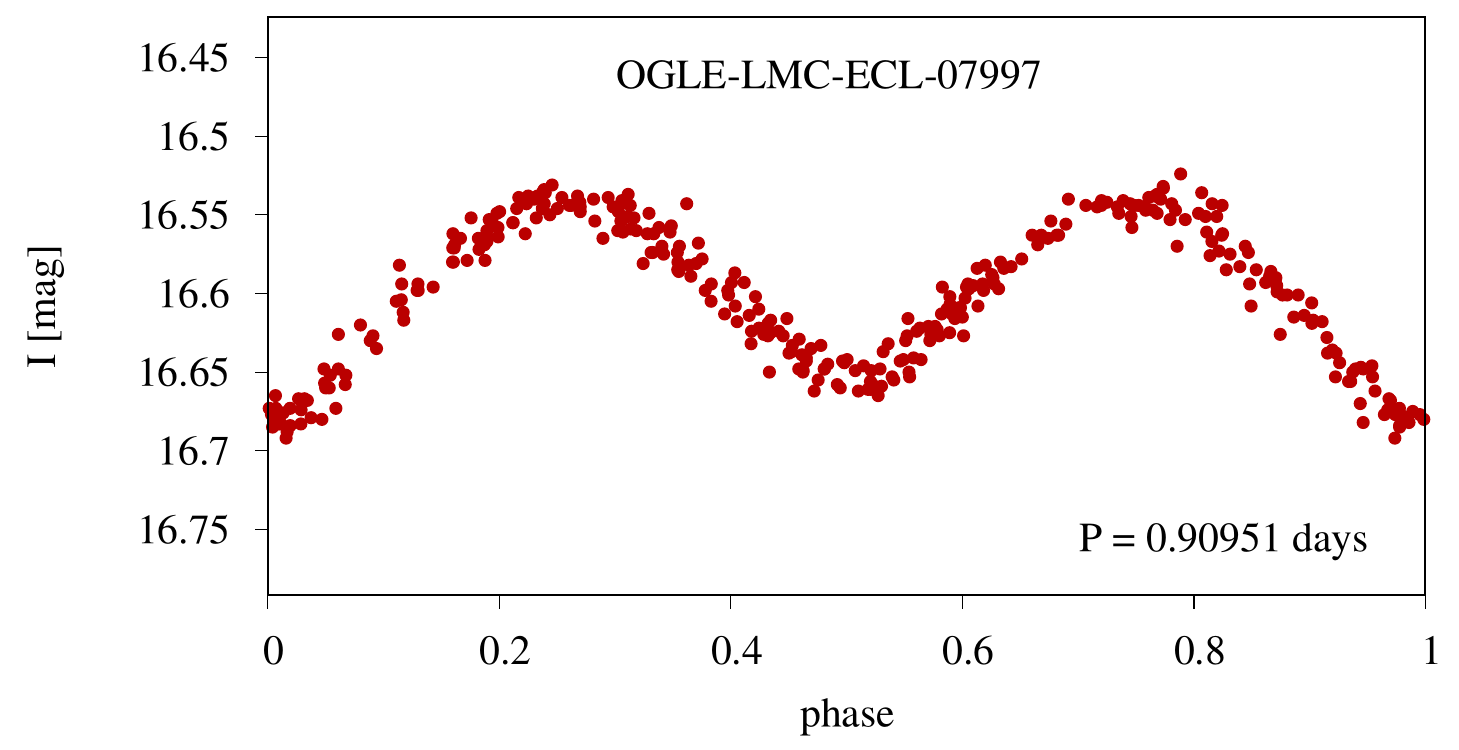}
    \includegraphics[clip=true, trim={0 0.5cm 0 0}, angle=270, width=0.32\linewidth]{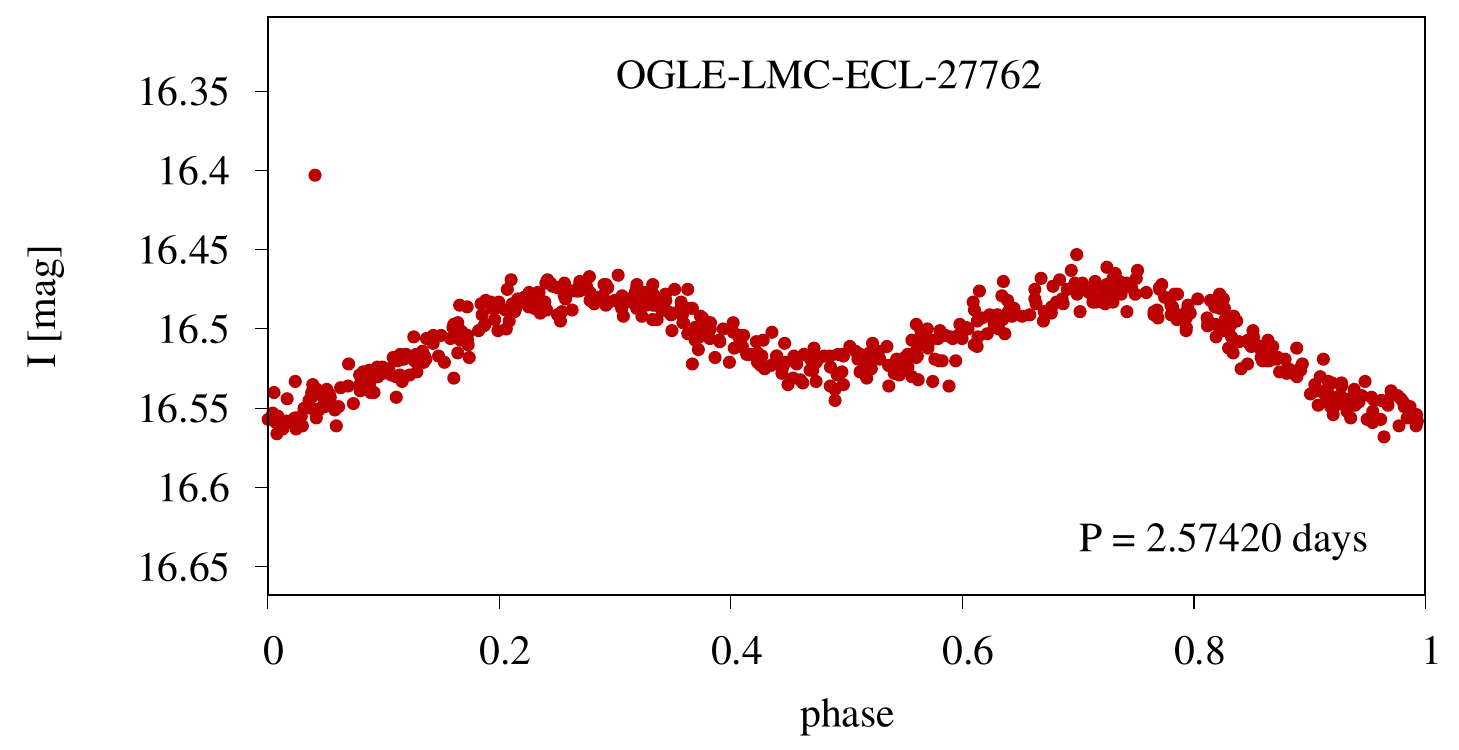}
    \end{subfigure}   
    \caption{Exemplary I-band light curves from the OGLE database,  classified by this study as EW (top), EB (middle) and ELL (bottom). The OGLE database itself uses a different classification scheme.}
    \label{OGLE_lc}
    \end{figure*}
    
We use the empirical period-luminosity-colour (PLC) relation for CBs from \citet{pawlak2016_plc}, to calculate the period of the widest O-type CBs. For an O4V-type MS CB, this is $\sim3$\,days. We note that there is significant scatter in this relation, therefore, this is only a rough estimate of the upper period limit for massive CBs.

Light curves are conventionally classified as: \textit{EW}- smooth sinusoidal light curves with  both eclipses having equal or very similar depths,  \textrm{EB}- light curves with barely-visible distinction between the eclipse and out-of-eclipse phase but unequal depths, \textrm{EA}- light curves with flat-bottomed minima and a clearer distinction between the eclipse and out-of eclipse phase and, \textrm{ELL} — smooth, sinusoidal light curves without eclipses, where only ellipsoidal modulation is visible. These typically arise from binaries viewed at low inclinations, making their intrinsic geometric configurations highly uncertain. Examples of EW, EB and ELL light curves from the OGLE database are shown in Fig.~\ref{OGLE_lc}.

In the OGLE database, however, light curves are not classified following these conventions. They were visually classified into three types: contact (C), non-contact (NC), and ellipsoidal (ELL), within our period range of interest. The label `C' indicates systems with EW light curves-- however, since these light curves were classified by eye, they were prone to errors. We therefore discard the OGLE labels, and systematically select binaries with light curves suited for our study.

Our primary assumption for selecting contact systems is that they are main-sequence binaries with EW-type light curves, wherein nearly-equal eclipse depths imply nearly equal-temperature components. This is because the eclipse-depth ratio is proportional to the fourth power of the temperature ratio of its components, $(T_2/T_1)^{4}$, or alternatively, their relative eclipse depth, $f_{\Delta A}\propto (T_2/T_1)^{4}$. We define the relative eclipse depth as, $f_{\Delta A} = \displaystyle \frac{(A_2 - A_1)}{((A_1+A_2)/2)}$, where $A_1$ and $A_2$ represent the difference between the  magnitude of the binary at maximum brightness, and the primary and secondary eclipse minima, respectively. Thus for the strict definition of EW light curves, $f_{\Delta A}\sim0$. However, since there are spectroscopic CBs reported with $(T_2/T_1)$ slightly away from 1, we relax this requirement for strictly equal minima  and include systems whose light curves show a nominal difference in eclipse depths, viz., with $f_{\Delta A}\leq0.1$. Within this constraint on eclipse depths, the radii ratio can be reasonably neglected. 

Next, we filter the sample based on their morph-parameter value provided by \citet{Bodi2021} for the above categories of light curves in the OGLE-IV catalogue. The morph parameter quantifies the geometric smoothness of the light curve shape and is a continuous value in the range $0 < c < 1$; the smoother the light curve is, the closer $c$ is to 1 \citet{Prsa2008}. \citet{Bodi2021} found that EW light curves have $c \gtrapprox 0.68$ (see their Fig.~1), thus setting a limit of $c\geq0.7$ on the OGLE sample leads to light curves with smooth sinusoidal shapes. 

To summarize, our empirical criteria for  the OGLE EW+ subsample which also includes binaries with slightly asymmetric light curves, are:

\begin{enumerate}
    \item They must have a high morph parameter ($c\geq0.7$).
    \item They must have orbital periods $\leq3$\,days, according to the empirical PLC of \citet{Pawlak2016}.
    \item Their absolute V-band magnitude is $M_\textrm{V,max}\lessapprox-2.5$ and absolute V-I colour is $-0.307\lessapprox M_\textrm{V,max}-M_\textrm{I,max}\lessapprox-0.165$.
    \item ELL binaries are excluded as their geometric configuration cannot be reliably determined.
    \item The relative eclipse depth between the primary and secondary components is $f_{\Delta A} \lessapprox0.1$. 
\end{enumerate}

Out of 2811 LMC and 861 SMC MS binaries, 56 systems in the LMC and 24 systems in the SMC satisfy these empirical criteria, and form our OGLE EW+ sample (the LMC sample is marked in Fig.~\ref{CMD_all}). Critically, these photometric–morphology criteria were defined prior to any reference to our synthetic CB predictions. This separation ensures that the empirically selected sample remains independent of theoretical expectations, which we describe in the next section.

\section{Synthetic light curves and populations of massive contact binaries}

\subsection{Binary evolution}
\label{models}
\begin{figure*}
\centering
\includegraphics[width=\linewidth]{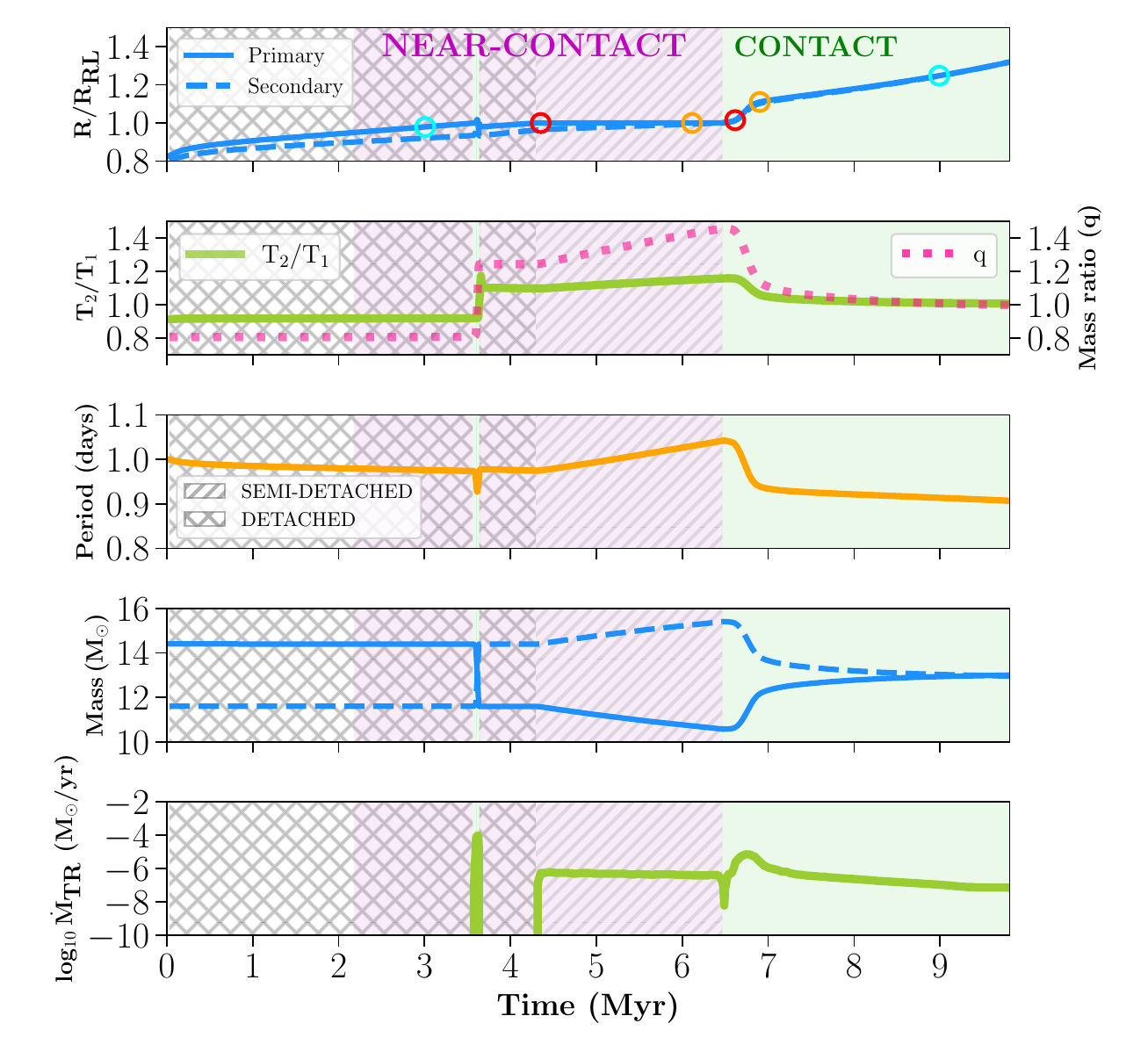}
    \caption{Panels showing the evolution of the exemplary `System 2' binary model from Paper I, with initial parameters: M$_\mathrm{T,i} = 26\,\mathrm{M_{\odot}}$, P$_\mathrm{i}$ = 1.0\,$\mathrm{d}$, q$_\mathrm{i}$ = 0.8. The parameters from top to bottom panels are: the fill-out factor (R/R$_\textrm{RL}$), the temperature and mass ratio (T$_2$/T$_1$; q), orbital period (P), mass of either component and the mass-transfer rate. Also shown are the contact (light-green region) phase and near-contact (pink region) phase, which is further divided into the detached phase (cross hatching) and semi-detached phase (striped hatching). The open circles are the instances at which synthetic \texttt{PHOEBE} light curves are computed  (Fig.~\ref{PHOEBE_LC}).}
    \label{orbital_fig}
\end{figure*}

\begin{figure*}
\centering
\includegraphics[width=\linewidth]{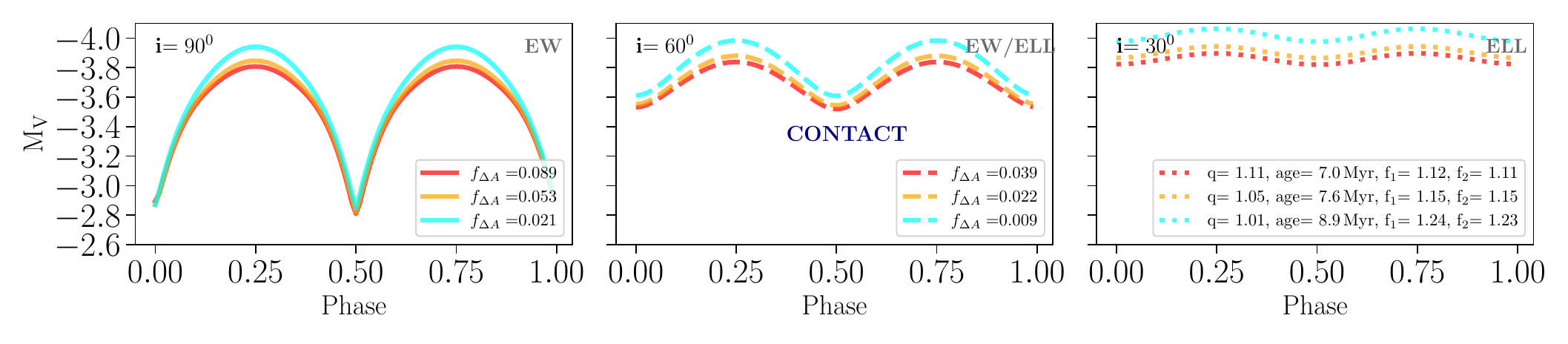}
\includegraphics[width=\linewidth]{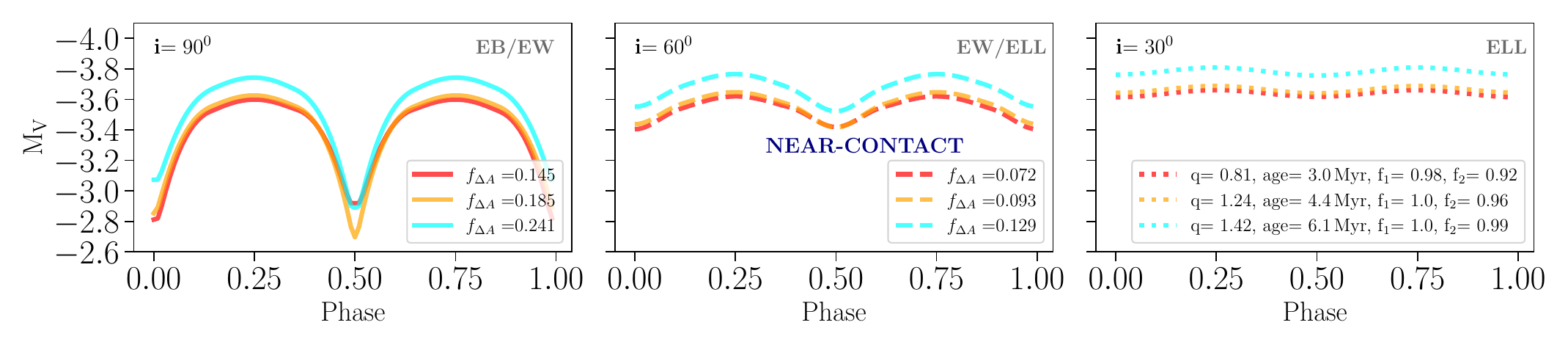}
    \caption{Light curves computed for System 2 at the various evolutionary points marked in Fig.~\ref{orbital_fig}, at inclinations of $i=30^{\circ},60^{\circ}$ and 90$^0$. Top panel: during the contact phase (green region in Fig.~\ref{orbital_fig}); the cyan light curve here represents the dominant type of CB expected from models, i.e., one which has acquired temperature equalization. Bottom panel: during the near-contact phase CB (pink striped region in Fig.~\ref{orbital_fig}), which is further split as a semi-detached (red and orange light curves ) and detached system (cyan light curve).  Also labelled are the relative eclipse depths, $f_{\Delta A}$, except for ELLs which have $f_{\Delta A}$ of the order of  0.001.}
    \label{PHOEBE_LC}
\end{figure*}

In our binary models, we define the contact and near-contact phases based on the Roche-lobe fill-out factors of the component stars, calculated as the ratio of their volume-equivalent radius to their Roche-lobe radius ($R/R_\text{RL}$). During the contact phase both stars have $R/R_\text{RL}>1$, and in the near-contact phase they have $R/R_\text{RL}
= 0.9-1.0$, following the same nomenclature as Paper I. 

In Paper I, two pathways to achieving nuclear-timescale contact were elucidated: the ``System 1" channel, where binaries achieved  early contact, just after the zero-age MS (ZAMS) and sustain contact until the binary overflows through the L2 Lagrangian point, and the ``System 2" channel, in which binaries acquired delayed contact after a semi-detached phase.  System 1 binaries dominate synthetic CB populations due to their significantly longer contact durations than System 2 binaries. However, System 2 binaries offer the opportunity to self-consistently study both the near-contact and contact phases, and examine the transition in light curve morphology between these phases.

 Fig.~\ref{orbital_fig} provides an overview of the evolution of the exemplary System 2 of Paper I,  with initial parameters, M$_\mathrm{T,i} = 26\,\mathrm{M_{\odot}}$, P$_\mathrm{i}$ = 1.0\,$\mathrm{d}$, q$_\mathrm{i}$ = 0.8, computed with \texttt{MESA} v.10398 \citep{paxton2011,paxton2013,paxton2015}. On achieving nuclear-timescale contact at $\approx6.4$\,Myr, this system evolves towards an equal-mass binary ($q\approx1$). Nuclear timescale contact results in temperature equalization following mass equalization, i.e., $T_2/T_1\approx1$ as $q\rightarrow1$, as seen in panel 2 of Fig.~\ref{orbital_fig}. 

 This model, like all others in this paper, does not include energy transfer within the shared envelope. Although the inclusion of envelope heat-transfer can increase the time spent in the $q<1$ phase during contact \citep{Fabry2023}, its overall impact on the parameter distribution of a population is minor \citep{fabry2025}. This will be discussed in Section~\ref{discussion}.

As the binary approaches contact, the system is first a  detached and later, a semi-detached system. The binary spends $\approx4.4$\,Myr in this near-CB phase, about 1.5 times longer than its contact phase, while maintaining a mass ratio $q < 1$, demonstrating that a mere 10\% difference in the Roche lobe fill-out factor results in a distinctly different structural configuration. As we will see in Section~\ref{comparison}, such near-contact binaries (near-CBs) are expected to be more numerous than CBs owing to their longer evolutionary timescales and typically have longer orbital periods, of $\approx1-2.5$\,days.

\subsection{Synthetic binary light curves}
\label{synthetic-lcs}

 We next investigate how the light curve of this binary behaves at different instances of its evolution at the  time stamps indicated in  Fig.~\ref{orbital_fig}. We compute synthetic light curves using PHOEBE II (hereafter PHOEBE; \citealp{Prsa2016, Horvat2018, Jones2020, Conroy2020}), which is based on the Wilson–Devinney code \citep{Wilson1971}. We assume a blackbody atmosphere for both stars, which provides a simplified but sufficient approximation for the purpose of light curve morphology. To compute the light curves, we supply the mass ratio (q), temperatures $T_1,T_2$, orbital period (P), inclination (i), primary mass and radius of the secondary, at each time-stamp of System 2 indicated in Fig.~\ref{orbital_fig}. 

Fig.~\ref{PHOEBE_LC} shows the resulting \texttt{PHOEBE} light curves. We find that the light curves are most sensitive to the inclination angle,  the temperature ratio and Roche-lobe fill-out factors ($R_1/R_\textrm{RL,1}, R_2/R_\textrm{RL,2}$). At high inclination angles, close to $90^0$, the configuration of the binary can be reliably inferred from the light curve shape with minimal uncertainty. 

 In the contact phase, the model has fill-out factors of $R/R_\textrm{RL}=1.1-1.2$ for both components and is predominantly a binary with $T_1\approx T_2$ and $q\approx1$. The corresponding EW \texttt{PHOEBE} light curves have relative eclipse depths of $f_{\Delta A} \approx 0.09-0.009$, as the inclination lowers from $90^{\circ}$ to $60^{\circ}$. Interestingly, the highest value of $f_{\Delta A}$ from our synthetic light curves (0.089) is close to the empirical upper limit of this parameter, while selecting  our OGLE EW+ subsample (0.10).

 In the phase prior to attaining  contact, the binary is a semi-detached or detached system binary nearing contact. The corresponding synthetic EB light curves have relative eclipse depths of $f_{\Delta A}=0.145-0.241$, as the mass ratio of the model in this phase is  $q\approx0.71-0.84$. These  EB light curves are visually similar to EW light curves except for their larger values of $f_{\Delta A}$, which may lead to their interpretation as belonging to unequal-mass CBs with $q<1$; examples of such systems may include V382 Cyg and TU Mus.

Distinguishing contact from near-contact binaries becomes even more challenging at lower inclinations. As the inclination angle approaches $60^\circ$, we find an increasing degeneracy between the light curves of the binary during the contact and near-contact phase. Even while the intrinsic $T_2/T_1\neq1$, the eclipse depth ratio approaches 1, or in other words, the relative eclipse depth, $f_{\Delta A}$ decreases up to $0.072$ at $60^{\circ}$. Thus as the inclination angle decreases, near-CBs can have EW/ELL-type light curves, which lends them the impression of being CBs with nearly-equal eclipse depths, although their intrinsic temperature ratio is $T_2/T_1<1$. At $i=30^{\circ}$ the situation worsens with an $f_{\Delta A}$ of an order of $0.001$, as only the ellipsoidal modulations of the binary are visible in the light curve, and it becomes exceptionally hard to discern the configuration of the binary. Thus a near-CB with a mass and temperature ratio $<1$, could be classified as having an EW/ELL light curve (Fig.~\ref{OGLE_lc}) at moderate-to-low inclinations.

Thus, while a 10\% in fill-out factors can result in a completely different evolutionary configuration of the binary, their  light curves, especially at  moderate to low inclinations, are not sensitive to this difference and are nearly indistinguishable between contact and near-contact binaries. To summarise:
\begin{enumerate}
    \item  Near CBs at moderate to low inclinations ($i\lessapprox60^{\circ}$) can exhibit light curves with nearly equal-eclipse depths, and get misclassified as CBs with EW light curves, particularly at $P\gtrsim1$\,days.
    \item Near CBs with smooth light curves that have sharp minima and unequal eclipse depths at high inclinations, could also be misinterpreted as CBs with $T_2<T_1$ and possibly, $q<1$.
    \item At extremely low inclinations ($i\lessapprox30^{\circ}$), only ellipsoidal variability is visible in the light curve (ELLs), making the intrinsic binary configuration difficult to determine. 
\end{enumerate}

These degeneracies between contact and near-contact binaries can be somewhat alleviated by studying synthetic populations and their parameter distributions.

\subsection{Binary-model grids and synthetic populations}
\label{numbers}

\begin{table*}
\caption{Parameters of the two grids of binary models used in this work-- The `Menon grid'  (\texttt{MESA} v. r10398, v. r15140) and the `Bonn grid' \citep{marchant_thesis} (\texttt{MESA} v. r8118). P$_\textrm{i}$ (days), M$_\textrm{1,i}$ (\Msun),  M$_\textrm{2,i}$ (\Msun), q$_\textrm{i}$ and $\alpha_\textrm{sc}$ are the initial orbital period, primary mass, secondary mass, mass ratio and semiconvection efficiency parameters  respectively.  f$_\textrm{contact/MS}$ and f$_\textrm{near-contact/MS}$ are the weighted fractions of CBs and near-CBs with respect to the MS populations in the individual grids,  respectively, along with W$_\textrm{grid}$/W$_\textrm{massive binaries}$, which assigns the relative contribution of each grid to the complete massive-star binary regime. These parameters are further explained in Section~\ref{numbers}.}
\centering
\begin{tabular}{ |c|c|c|c|c|c|c|c|c|c| } 
\hline
Grid & P$_\textrm{i}$ (days) & M$_\textrm{1,i}$ (\Msun) & M$_\textrm{2,i}$ (\Msun) &  q$_\textrm{i}$ & $\alpha_\textrm{sc}$ & f$_\textrm{contact/MS}$ & f$_\textrm{near-contact/MS}$ & W$_\textrm{grid}$/W$_\textrm{massive-binaries}$\\
\hline
Menon  & 0.6...2 & 7...50 & 7...40 & 0.6...1.0 & 1.0 & 0.1863 & 0.1355 & 0.075\\ 
Bonn & 2.1...3162 & 10...40 & 7...40 &  0.275...0.975 & 0.01 & 0.036 & 0.0114 & 0.450\\
\hline
\end{tabular}
\label{grid}
\end{table*}

We combined two grids of models in this work: the `Menon grid' from Paper I for binaries with initial periods of $0.6-2$\,days and the `Bonn grid' \citep{marchant_thesis,Sen2022}, spanning $2.1-3162$\,days using  \texttt{MESA} \citep{paxton2011,paxton2013, paxton2015,paxton2018,paxton2019}. We extend the Menon grid to ensure full coverage of B-type  binaries, by computing new models with initial masses of $M_\textrm{T,i}=14...18$\Msun. Table~\ref{grid} summarizes the parameters and relative weights of each grid. These parameters reflect the parameter space coverage for massive binaries, based on the findings of \citet{sana2013} of 30~Doradus, from the VLT-FLAMES Tarantula survey \citep{evans2011}. Thereafter, using the same assumptions and methods as Paper I, we construct synthetic populations from this combined model grid and compute parametric probability distributions (further described in Section~\ref{populations}). 


Paper I had also shown that the distributions of binary properties are similar for the metallicities of the  LMC and SMC for our systems of interest. In Section~\ref{MS_sample}, we also found that the binary fraction at different luminosities is similar between the LMC and SMC. Therefore, in this work, we construct the probability distributions only for the LMC and adapt it to the SMC, by scaling it to the number of binaries in the SMC. Case-A mass transfer is expected to occur in binaries with initial orbital periods up to $P_\mathrm{i} \approx 23$\,days, which defines the upper limit for systems contributing to the contact and near-contact binary populations. The number of systems calculated from our synthetic populations (the details of which are in Section~\ref{populations}) are:\\

$N_\textrm{contact, LMC}\approx 39,  N_\textrm{near-contact, LMC}\approx 73$\\
\indent$N_\textrm{contact, SMC} \approx 12,  N_\textrm{near-contact,  SMC} \approx 23$ \\

Near-CBs are predicted to be nearly twice as common as contact systems in a population, making them more likely to be observed. The predicted counts of 39 CBs in the LMC and 12 in the SMC serve as benchmarks for comparison with the OGLE sample.

We note that the contribution to the CB population solely originates from the Menon grid, i.e., from binaries with initial periods under 2\,days. Including the Bonn grid effectively broadens the parameter space over which near-CBs may be observed compared to Paper I, as we discuss in the next section.

\section{Comparing massive CB candidates with theoretical populations}
\label{comparison}

We investigate synthetic distributions for three properties from the OGLE catalogue: the absolute magnitude ($M_\textrm{V,max}$), colour ($M_\textrm{V,max}-M_\textrm{I,max}$), and orbital period ($P$).

\texttt{MESA} models provide the intrinsic luminosities of the binary components, whose sum corresponds observationally to the maximum absolute brightness of the system at an inclination of $i = 90^\circ$. We first convert the bolometric luminosities of the primary and secondary stars in each binary model to their respective absolute V- and I-band magnitudes using the flux–magnitude relation and bolometric corrections from  Section~\ref{MS_sample}. We then reconvert the V- and I-band magnitudes of each stellar component into their intrinsic luminosities, compute the total binary luminosity in each band ($L_{\mathrm{binary,band}} = L_{1,\mathrm{band}} + L_{2,\mathrm{band}}$), and finally convert $L_{\mathrm{binary,band}}$ to the absolute V- and I-band magnitudes of the binary ($M_\textrm{V,max}$ and $M_\textrm{I,max}$) using the flux–magnitude relation once more.

\begin{figure*}
   \centering
    \includegraphics[width=0.45\linewidth]{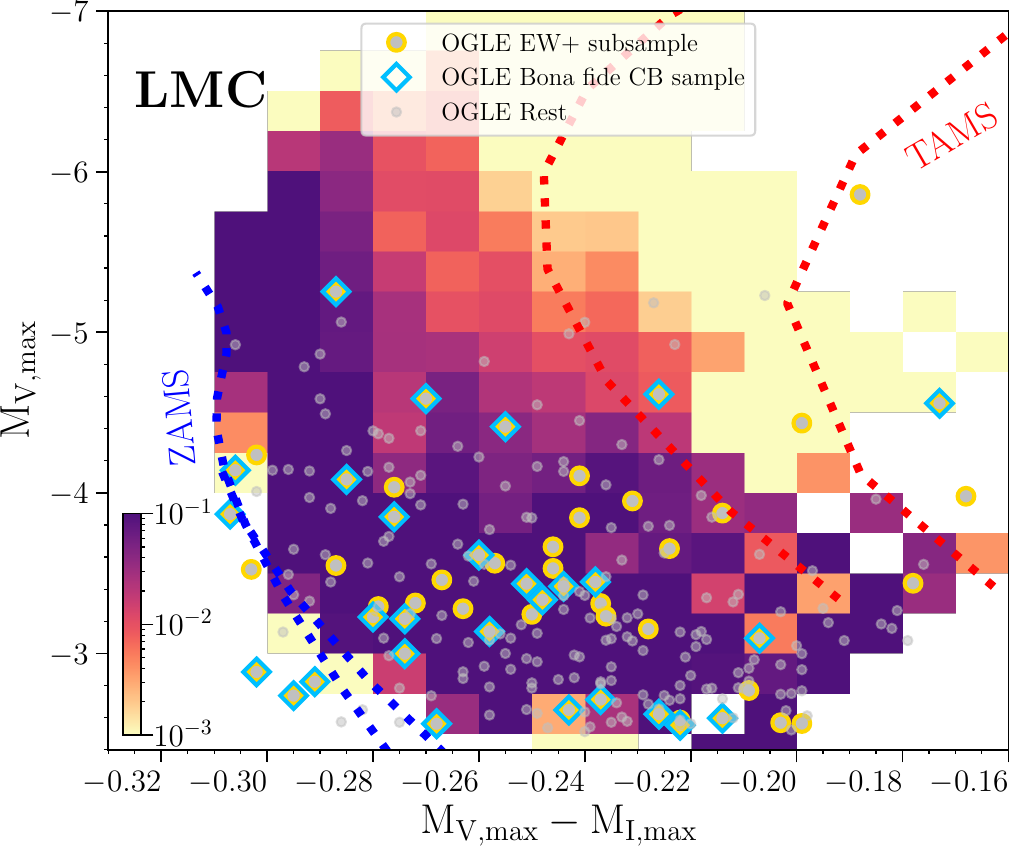}
    \includegraphics[width=0.45\linewidth]{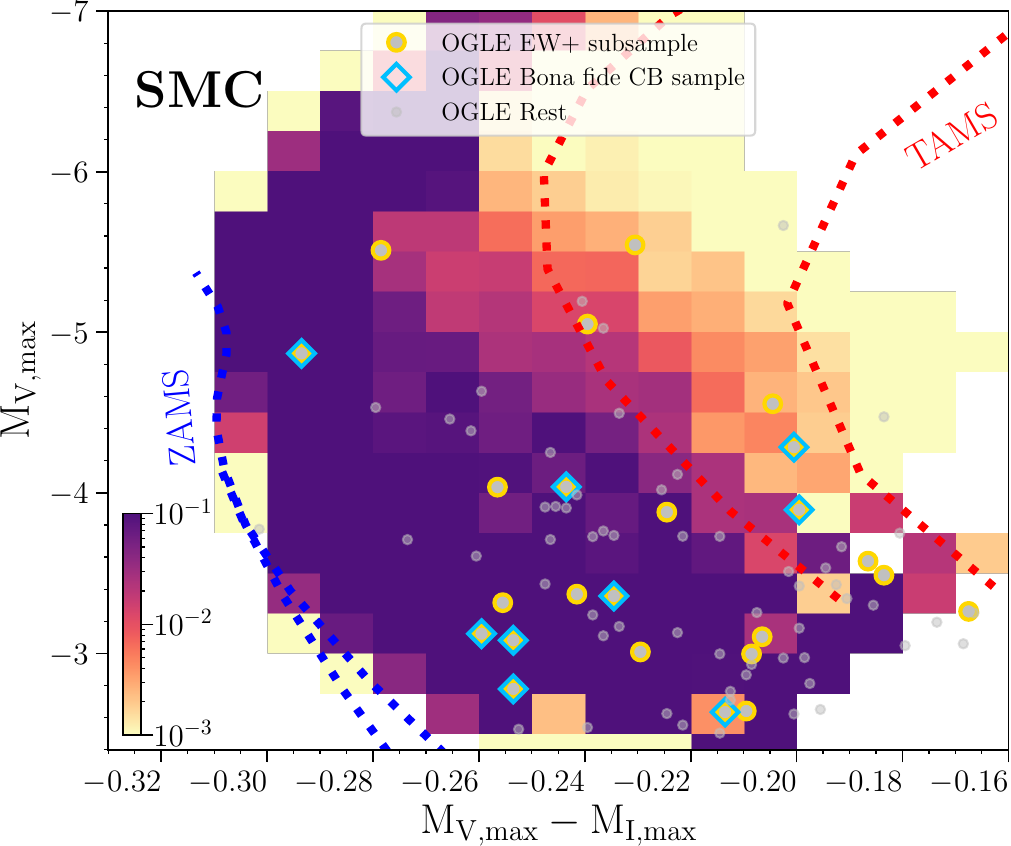}    
    \caption{Heat map showing the probabilistic distribution of the absolute colour (M$_\mathrm{V,max}$-M$_\mathrm{I,max}$) and magnitude (M$_\mathrm{V,max}$) at the maximum brightness (i.e., the sum of intrinsic luminosities of both components) of CBs, for the LMC and SMC. OGLE data is marked as: OGLE Bona fide CB subsample (blue diamonds),  OGLE EW+ subsample (yellow circles) and OGLE Rest (grey dots), which is the sample with morph parameter $c>0.7$ and $P>3$\,days that did not make the cut into either the EW+ or the Bona Fide CB subsample. For reference, the theoretical ZAMS and TAMS curves from Fig.~\ref{CMD_all} are also marked.}
              \label{CMD}%
    \end{figure*}

Fig.~\ref{CMD} shows the location of our OGLE EW+ sample on the theoretical colour-magnitude distribution (CMD). The B-type CBs appear more evolved than the O-type CBs—likely because their longer contact durations allow them to remain on the main sequence longer before merging. A few OGLE systems are bluer than our theoretical ZAMS curves, suggesting an overestimated extinction (as noted in Section~\ref{MS_sample}). Overall, the locations of our mined OGLE sample agree well with the theoretical CMD for massive CBs.

\begin{figure*}
   \centering
 \includegraphics[width=0.49\textwidth]{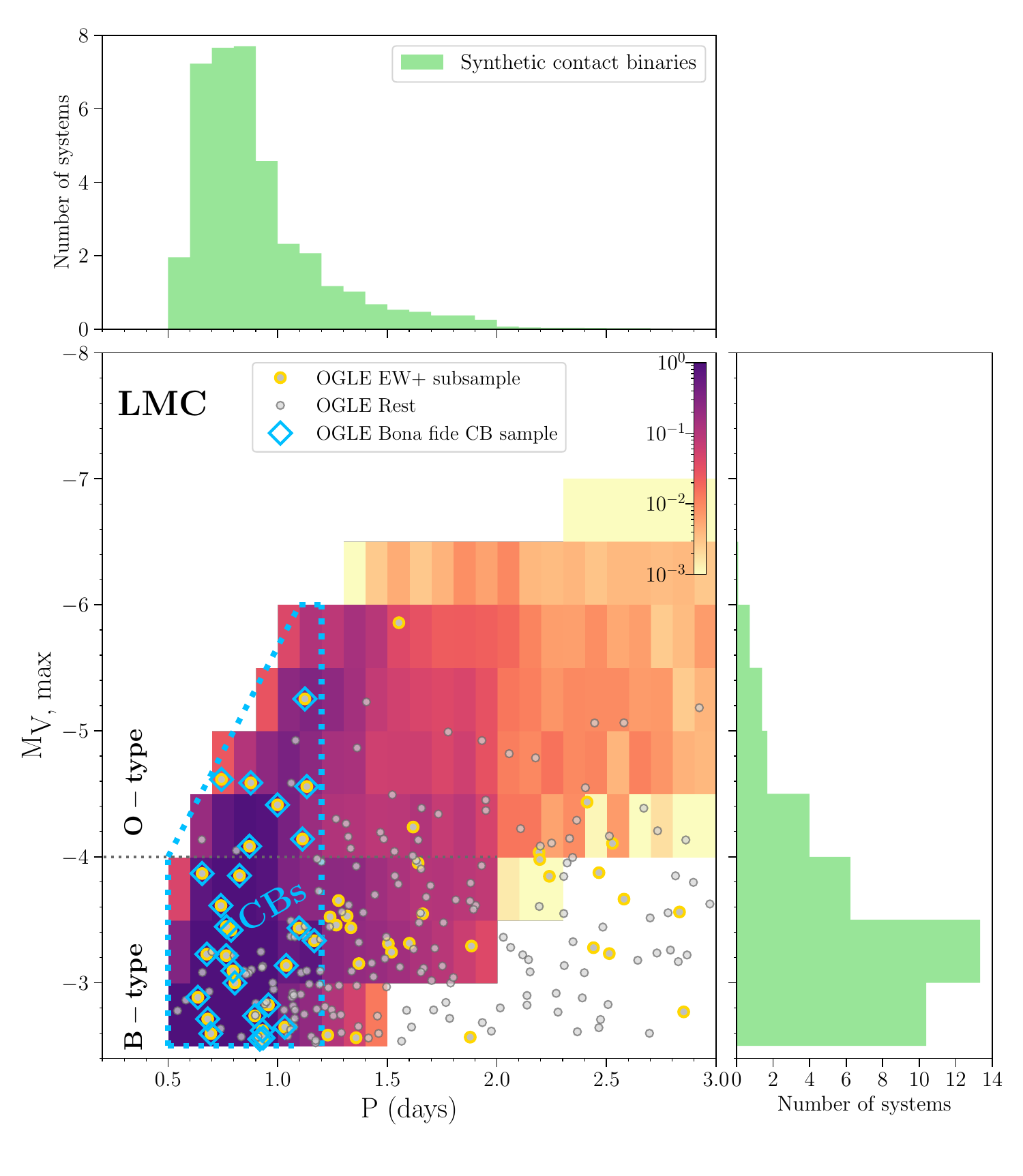}
 \includegraphics[width=0.49\textwidth]{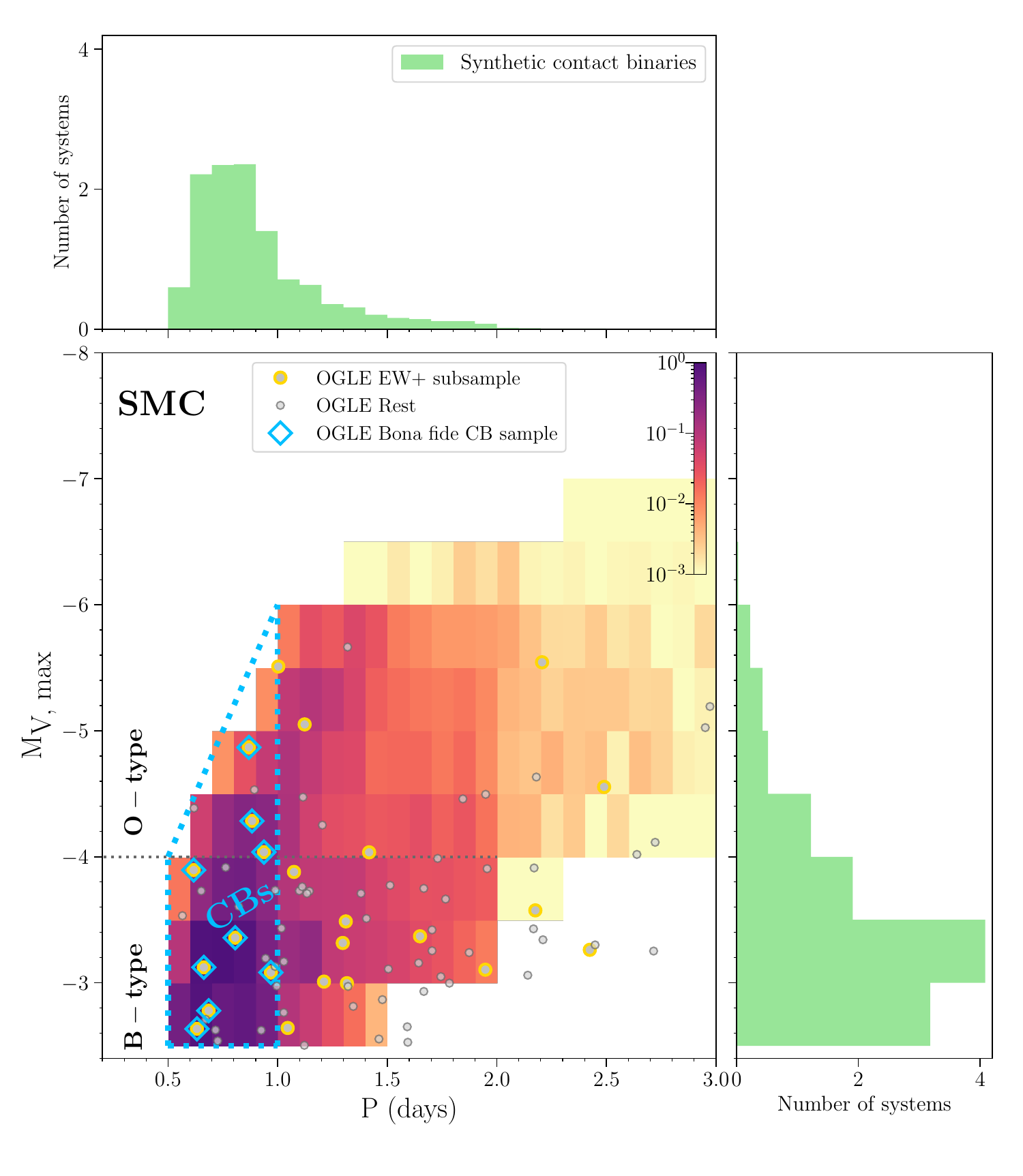}
    \caption{Heat map showing the probabilistic distribution of the absolute magnitude (M$_\mathrm{V,max}$) at maximum light to orbital period (P) of CBs, for the LMC and SMC. O and B-type systems are distinguished by a magnitude cut-off at M$_\mathrm{V,max}=-4$. OGLE data is annotated similar to Fig.~\ref{CMD}. For reference, the theoretical extent of the Bona Fide CB subsample is marked as the blue dotted blue polygon, beyond which their number counts from de-convoluted histograms (green histograms) drop below 1.}
                 \label{Mv-P-contact}%
    \end{figure*}

Fig.~\ref{Mv-P-contact} shows the probability distributions for CBs from our binary-model grid in the period–magnitude plane, together with our empirical OGLE EW+ sample (yellow circles), and all other high-morph parameter binaries occupying the same period–colour–magnitude domain but with larger relative eclipse depths  $f_\Delta A>0.1$ (grey dots). Although the empirical PLC extends to \(P\approx3\)\,days, our OGLE EW+ sample shows clear outliers beyond 2\,days, deviating from the expected trend and suggesting that the PLC relation may not reliably apply to massive systems in this period regime.

The high‐likelihood region for CBs, indicated by the blue polygon in Fig.~\ref{Mv-P-contact}, extends to \(P\approx1.2\)\,days in the LMC and \(P\approx1.0\)\,days in the SMC. These represent the widest CBs expected from our models, beyond this limit the population count for CBs is smaller than 1.  Systems falling within this region form our ``bona fide CB sample'. These comprise of 28 CB candidates in the LMC and 9 candidates in the SMC. Of these 37 systems, 27 are B-type binaries with orbital periods of 0.6--1.0\,days. This sample also includes the two known spectroscopic CBs in the Clouds--  VFTS 352 \citep{almeida2015} in the LMC and OGLE SMC-SC10 108086 system \citep{hilditch2005}. The full set of bona fide CB candidates and the remainder of the EW+ sample are listed in Tables~\ref {LMC_OGLE_CBs} and \ref{SMC_OGLE_CBs}.

For comparison with the theoretical population counts (Section~\ref{numbers}), we also consider the broader census of EW+ binaries (including the bona fide sample) and ELLs  within the high-likelihood region of CBs, yielding: 

\begin{align*}
N_{\rm OGLE\text{-}bona\ fide,\ LMC} &= 28, \quad 
N_{\rm OGLE\text{-}bona\ fide,\ SMC} = 9 \\
N_{\rm OGLE\text{-}ELL,\,P\leq1.2\,{\rm days},\ LMC} &= 12, \quad 
N_{\rm OGLE\text{-}ELL,\,P\leq1\,{\rm days},\ SMC} = 3 \\
N_{\rm OGLE\text{-}total,\,P\leq1.2\,{\rm days},\ LMC} &= 40, \quad 
N_{\rm OGLE\text{-}total,\,P\leq1\,{\rm days},\ SMC} = 12
\end{align*}

These numbers are remarkably close to our theoretical estimates from Section~\ref{numbers}, indicating a consistency between the data and theory. The period, magnitude and colour distributions of the bona fide sample shown in Figs.~\ref{Mv-P-contact} and Fig.~\ref{P-q} are consistent with the CB population predicted from our MESA models. Our prediction from Fig.~\ref{P-q}  for this mined sample is that these binaries are likely to have mass ratios of $q\approx1$. Further spectroscopic follow-up with RV data is required to confirm our prediction

Fig.~\ref{P-q} also shows that near-CBs are expected to be twice as numerous as CBs and have orbital periods of $\sim1-2.5$\,days and mass ratios $q\approx0.4-1$. This orbital period limit is longer than the number predicted in Paper I, due to the inclusion of the Bonn grid in our work. A fraction of these systems may have light curves similar to CBs, as demonstrated in Section~\ref{synthetic-lcs}, and masquerade as CBs with $P\gtrapprox1$\,days in a population.

    
\section{Discussion}
\label{discussion}

We have empirically mined 56 LMC and 24 SMC main-sequence binaries from the OGLE survey with EW light curves that may belong to stellar components with nearly-equal temperatures. Of these, a total of 37 systems from both Clouds form our bona fide contact binary (CB) candidate sample, which is the largest homogeneous sample of massive CB candidates in literature. 
Including ELL binaries further increases  this sample, and brings the sample size strikingly close to our predicted number of CBs-- 39 and 12 systems in the LMC and SMC respectively. Accounting for the OGLE survey’s $\sim$80\% classifier recall suggests that the underlying population of massive CBs may be even larger. Conversely, incorporating the detailed star‐formation histories of the Magellanic Clouds could reduce these estimates.  During the preparation of this manuscript, OGLE released an updated database \citep{glowacki2024}, the consideration of which may further affect the above numbers.

Our models suggest that the bona fide CB candidates are likely undergoing a nuclear-timescale contact phase, during which mass equalization leads to subsequent temperature equalization. According to our synthetic populations, such $q\approx1$ CBs are expected to dominate the overall census of massive CBs. This result is further reinforced by the recent work of \citet{fabry2025}, who included additional physics such as envelope energy transport and tidal deformation in their CB models (which were not included in our models).  By computing several models, they inferred that while heat-transfer physics allows for a minority population ($\lessapprox 10\%$) of delayed-contact CBs with $q<1$, the dominant majority of CBs still remained B-type binaries that attained contact soon after ZAMS. Therefore, even with the inclusion of these physics mechanisms, the $P-q$ distribution of massive CBs remained largely dominated by $q\approx1$ systems with periods less than 1\,day, consistent with our predictions. 

Such equal-mass CBs are not uncommon in literature. Of the two previously reported spectroscopic CBs in the Magellanic Clouds, the deep-contact binary VFTS 352 \citep{almeida2015, mahy2020b}, which is also part of our OGLE sample, has a mass ratio of $q\approx1$.  The sole spectroscopic CB reported in the SMC, OGLE SMC SC10-108086 \citep{hilditch2005}, is also part of our sample. Its reported mass ratio of $q = 0.85^{+0.056}_{-0.056}$, suggests that this system can have a mass ratio up to $q = 0.91$, which although not exactly  equal to one, is close to a mass ratio of unity. B-type spectroscopic binaries in deep contact in the MW also have equal-mass components, as do two recent photometric systems discovered in the M31 galaxy \citep{Li2022}. VFTS 066 was earlier classified as a CB in \citet{mahy2020b}, however, this system has an inclination of $i=17.5^{\circ}$ and only shows ellipsoidal variability \citep{vrancken2024}. Thus its intrinsic configuration cannot be definitively ascertained.
 
 A close examination of the literature reporting  $q<1$ CBs, reveals that the uncertainty in their Roche-lobe fill-out factor allows for their interpretation as binaries only approaching contact. Examples of such systems include V382 Cyg \citep{yasaroy2013, martins2017}, TU Mus \citep{penny2008} and LSS 3074 \citep{raucq2017}, which have been described as "marginally in contact", ``shallow contact'' or ``approaching contact'', highlighting the ambiguity in the configuration of reported contact systems. By analysing spectroscopic and photometric data, V729 Cyg \citep{yasaroy2014} and V382 Cyg \citep{yasaroy2013, martins2017} can be classified either as a binary in ``weak'' contact, barely filling its Roche lobes or a semi-detached binary nearing contact; V729 Cyg may in fact not even be a binary on the main-sequence, but may have a component which is a Wolf-Rayet star. These uncertainties in fill-out factors, coupled with the presence of such atypical systems, suggest that previously known spectroscopic samples likely represent a mixed population of contact and near-contact binaries, with the latter expected to occur at roughly twice the frequency of the former.

At this juncture, we wish to emphasize that the compendium of spectroscopic systems presented in Paper I represents a mixed population, with many \( q < 1 \) systems occupying ambiguous contact or near-contact configurations. Indeed, our PHOEBE models demonstrate that contact and near-contact configurations can be degenerate, implying that a fraction of near-CBs with smooth EW light curves could be mistaken for contact systems, especially at low inclinations. We encourage the reader to carefully assess whether a given system in the sample compendium of Paper I may be interpreted as a \( q < 1 \) semi-detached or detached binary, and to avoid overstating the likelihood of  genuine CBs with \( q < 1 \) CBs.

On that note that we do not preclude the possibility of $q<1$ massive CBs, nor make predictions for their population counts, which is beyond the scope of this study. A fraction of these systems may be included in our sample, since we also accommodate systems with slightly unequal temperatures, and therefore possibly, unequal masses. We also do not rule out the existence of CBs with periods longer than 1.2\,days, which are included in the `extended sample' of systems in Tables~\ref{LMC_OGLE_CBs} and \ref{SMC_OGLE_CBs}.  A multi-epoch spectroscopy campaign is essential to determine the mass ratios of our sample and the true numbers of CBs.

\section{Conclusions}
\label{conclusions}

We have identified a total of 37 bona fide CB candidates from both Magellanic Clouds, with nearly equal eclipse depths, which maybe associated with binaries containing equal-temperature components. Notably, this sample also contains 27 B-type systems, thus  addressing the long-standing dearth of B-type CBs and equal-temperature CBs.  Along with 10 O-type systems (of which two were previously known), we have increased the number of massive CB candidates in the Magellanic Clouds, by an order of magnitude. Incorporating additional ELL-type binaries brings the sample into close agreement with our theoretical predictions. However, classification uncertainties and the underlying star-formation histories can also affect the true population size of CBs.

Our theoretical models and synthetic light curves suggest that these CB candidates may have undergone mass equalization prior to achieving temperature equalization. Therefore, we expect a subset of the sample to represent the previously under-reported population of $q \approx 1$ massive CBs. 

The synthetic light curves we constructed for evolutionary models of semi-detached and detached systems nearing contact were found to be similar to CBs, except for their distinctly unequal eclipse depths. This morphological similarity may lead to the misidentification of some near-CBs as CBs with highly unequal mass components, which becomes more pronounced  at low inclinations. Thus, a 10\% difference in the Roche-lobe fill-out factor, which appears modest from an observational perspective, can mask critical differences in physical configuration and result in observational ambiguity between contact and near-contact states. 

Our theoretical predictions and candidate CB sample suggest that massive contact binaries comprise about 1-2\% of the massive main-sequence binary population. Despite their small numbers, these systems represent an important population, serving as an observational portal into the under-studied channel of binary evolution involving stellar mergers. They also provide a crucial testing ground for confronting binary evolution theory with empirical data. Our study lays out a promising pipeline bridging evolutionary models and photometric data. Multi-epoch radial-velocity data from dedicated spectroscopic studies or as part of surveys like the ongoing BLOeM survey \citep{Shenar2024} and the upcoming 4MOST survey \citep{Cioni2019,deJong2019}, combined with deep learning techniques and machine-learning classifications, will further develop this pipeline and  enable full-fledged population studies of various binary classes and the distribution of their properties. 

\begin{acknowledgements}
AM was supported by the Juan de la Cierva Incorporación fellowship IJC2020-045628-I for this work. MP is supported by the BEKKER fellowship BPN/BEK/2022/1/00106 from the Polish National Agency for Academic Exchange and the Royal Physiographic Society in Lund through the Märta and Erik Holmbergs Endowment. We thank Michael Abdul-Masih for their help with \texttt{PHOEBE} and Artemio Herrero for insightful discussions that helped shape this paper.

\end{acknowledgements}

%
   \bibliographystyle{aa} 
   \bibliography{master.bib} 
%
\newpage

\begin{appendix}

\section{Construction of synthetic populations}
\label{populations}
  \begin{figure*}
   \centering
    \includegraphics[width=0.48\linewidth]{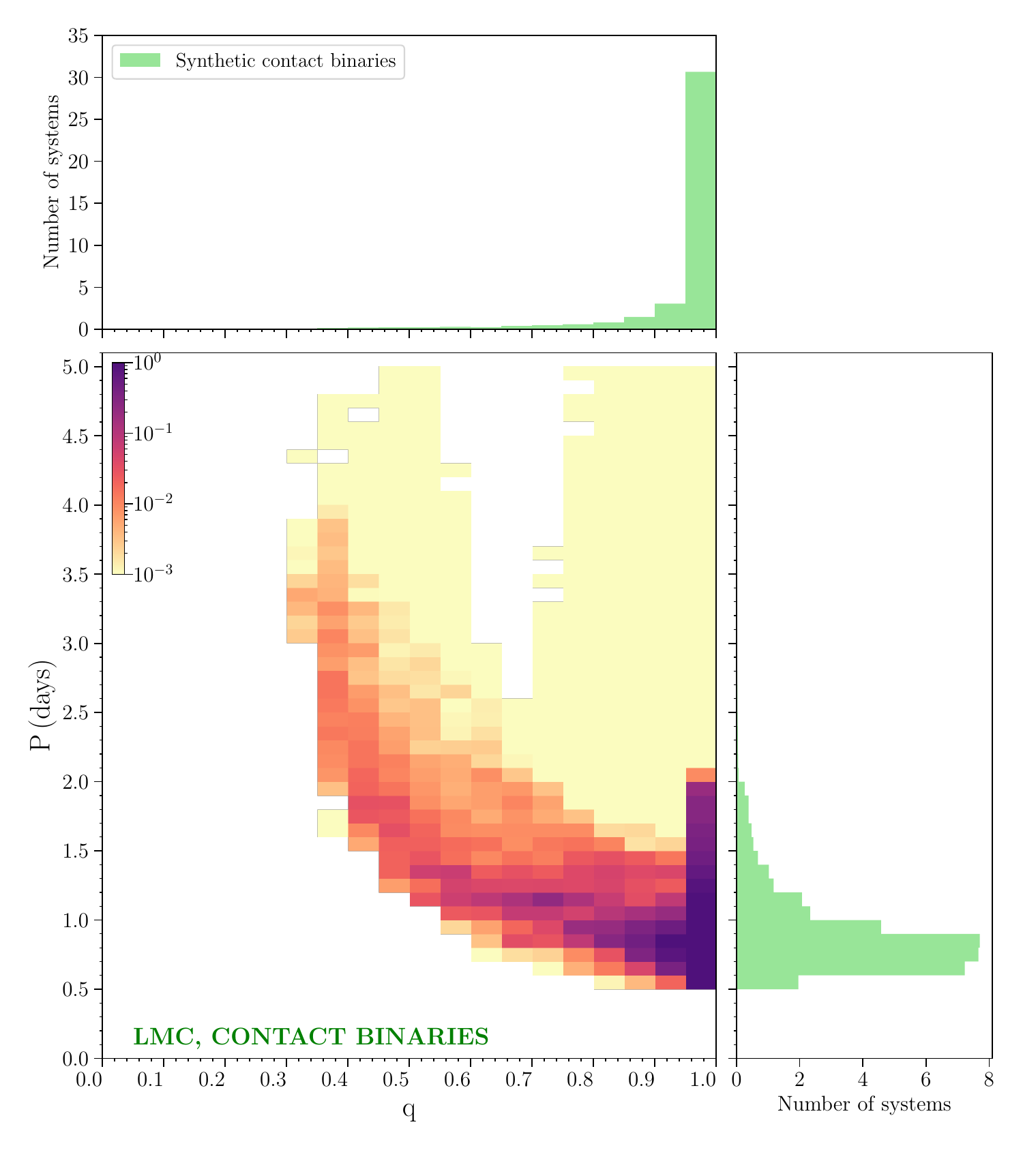}
\includegraphics[width=0.48\linewidth]{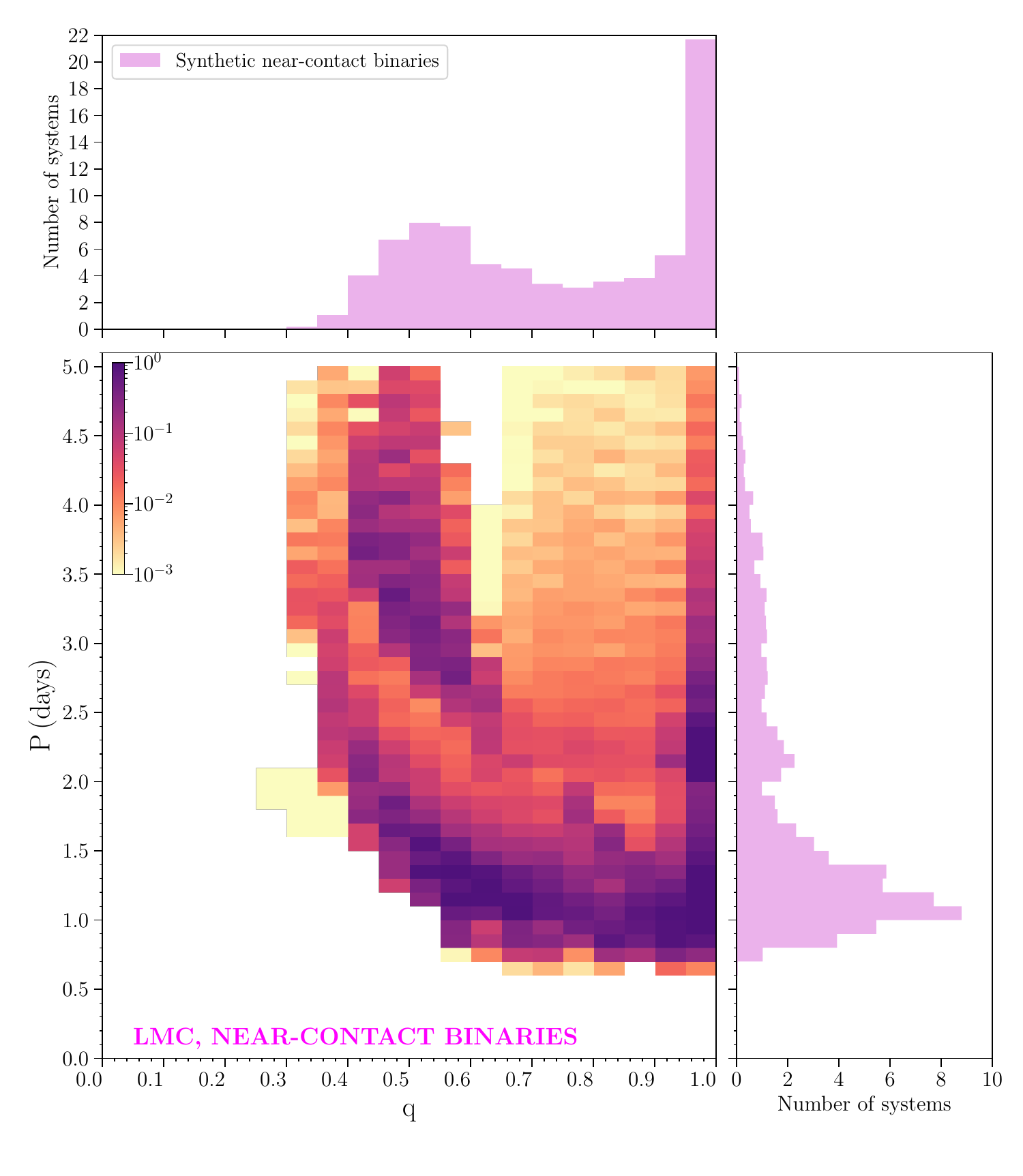}
   \caption{Heat map showing the probability distribution of period (P) and mass ratio (q) from our LMC synthetic population of CBs and near-CBs, tailored for the OGLE sample. The SMC has a closely similar $P-q$ distribution as the LMC for massive CBs, as determined in Paper~I.}
              \label{P-q}%
    \end{figure*}
    
We assume a constant star formation rate and adopt a Salpeter initial mass function (IMF) for primary masses, along with flat distributions in log-period and mass ratio. For the initial parameter space encompassing all massive binaries, we consider the parameter limits sampled by the VLT-FLAMES \citep{sana2013,dunstall2015} and Tarantula Massive Binary Monitoring (TMBM) \citep{almeida2015} surveys of 30~Doradus, which are: initial primary masses \(6 \le M_1/M_\odot \le 90\), mass ratios \(0.275 \le q \le 1.0\), and orbital periods \(0.6 \le P_i\,\mathrm{d} \le 3162\). We denote the `bulk weight' corresponding to this initial parameter space of all massive binaries as  W$_\textrm{massive binaries}$. Similarly, we determine the bulk weights of either model grid (Menon or Bonn) based on their initial parameter range:

\begin{align*}
W_\textrm{massive binaries} = |(1-0.275)\cdot(\textrm{log}_{10}(3162)-\textrm{log}_{10}(0.6))\cdot\\
(90^{-1.35} - 6^{-1.35})| = 0.023\\
W_\textrm{Menon-grid} = |(1-0.6)\cdot(\textrm{log}_{10}(2)-\textrm{log}_{10}(0.6))\cdot\\
(50^{-1.35} - 6^{-1.35})| = 0.017\\
W_\textrm{Bonn-grid} = |(0.975-0.275)\cdot(\textrm{log}_{10}(3162)-\textrm{log}_{10}(2))\cdot\\
(40^{-1.35} - 10^{-1.35})| = 0.084\\
\end{align*}

The coverage fraction of either grid, relative to the complete parameter space of massive binaries is: 8\% for the Menon grid and 45\% for the Bonn grid. The reason the sum of these fractions is not 1, is because neither grid extends to the complete mass and period range of massive stars. Regardless, our grids sufficiently cover the parameter space relevant for short-period MS systems that are studied in this work. 

To determine the  probability distributions of CBs and near-CBs from our synthetic populations, we revisit the method of Paper I. We begin by calculating the  fraction of the CB and near-CB population to the MS binary population in each grid, according to equation~8 in  Paper I: 

\begin{align}
\mathbf{f}_{\mathrm{contact}/\mathrm{MS}} = 
\frac{\sum_{s=1}^{n} w_s\, \tau_{\mathrm{contact},s}} 
     {\sum_{s=1}^{n} w_s\, \tau_{\mathrm{MS\text{-}binary},s}} \,.
\label{eq:f_contact_MS}
\end{align}

where, s is an index that loops through all binary models in each grid, $w_\textrm{s}$ stands for the birth weight of the system, using the priors on  IMF, period and mass ratio discussed at the beginning of this section, $\tau_\textrm{contact,s}$ and $\tau_\textrm{MS-binary,s}$ are the contact duration and MS lifetimes, respectively. The denominator of Eq.~\ref{eq:f_contact_MS} considers the MS lifetime of each binary across the entire grid, so that $f_\textrm{contact/MS}$ ($f_\textrm{near-contact/MS}$) gives us the overall fraction of CBs (near-CBs) in a population of MS binaries. 

We then compute the expected number of contact and near-contact systems in each galaxy as: 
\begin{align}
\label{N_eq1}
N_\mathrm{contact,\ grid} &= 
\frac{W_\mathrm{grid}}{W_\mathrm{massive-binaries}} \cdot 
f_{\frac{\mathrm{contact}}{\mathrm{MS}}}\cdot 
N_\mathrm{OGLE,\ MS} \, \\
\label{N_eq2}
N_\mathrm{near\text{-}contact,\ grid} &= 
\frac{W_\mathrm{grid}}{W_\mathrm{massive-binaries}} \cdot 
f_{\frac{\mathrm{near-contact}}{\mathrm{MS}}}\cdot 
N_\mathrm{OGLE,\ MS} \,
\end{align}

where, $\displaystyle \frac{W_\textrm{grid}}{W_\textrm{massive -binaries}}$ stands for the coverage in initial parameter space by each grid  as a  fraction of the total initial parameter range for all massive binaries in the Universe. 

N$_\textrm{OGLE, MS}$ stands for the number of OGLE MS binaries (as described in Section~\ref{MS_sample}),  $f_\textrm{contact/MS}$ and $f_\textrm{near-contact/MS}$ are the weighted fraction of contact and near-contact binaries from either grid, computed according to Eq.~8 of Paper I.

Plugging in $N_\textrm{OGLE, MS}=2811$ for the \textbf{LMC} in equations~\ref{N_eq1} and \ref{N_eq2}, with values from Table~\ref{grid}, we get:
\begin{align*}
N_\mathrm{contact,\ Menon\text{-}grid} &= 39, \quad 
N_\mathrm{contact,\ Bonn\text{-}grid} = 0 \,, \\
N_\mathrm{near\text{-}contact,\ Menon\text{-}grid} &= 28, \quad 
N_\mathrm{near\text{-}contact,\ Bonn\text{-}grid} = 45 \,, \\
N_\mathrm{contact,\ LMC} &\approx 39, \quad 
N_\mathrm{near\text{-}contact,\ LMC} \approx 73, \quad 
\end{align*}
\vspace{-20pt}
\begin{center}
$N_\mathrm{combined,\ LMC} \approx 113$
\end{center} 

Using $N_\textrm{OGLE,massive-MS}=861$ for the \textbf{SMC}, we get:
\begin{align*}
N_\mathrm{contact,\ Menon\text{-}grid} &= 12, \quad 
N_\mathrm{contact,\ Bonn\text{-}grid} = 0 \,, \nonumber \\
N_\mathrm{near\text{-}contact,\ Menon\text{-}grid} &= 9, \quad 
N_\mathrm{near\text{-}contact,\ Bonn\text{-}grid} = 14 \,, \nonumber \\
N_\mathrm{contact,\ SMC} &\approx 12, \quad 
N_\mathrm{near\text{-}contact,\ SMC} \approx 23 \,, \nonumber \\
\end{align*}
\vspace{-30pt}
\begin{center}
$N_\mathrm{combined,\ SMC} \approx 34$
\end{center} 

To obtain the number distributions shown in Figs.~\ref{CMD} to \ref{P-q}, we multiply the p.d.fs with the above numbers for each grid, and  combine the total weighted population across both grids to a seamless distribution function for each property considered.

\section{OGLE CB candidates in the LMC and SMC}
\label{OGLE-tables}

\begin{table*}
\centering
\caption{LMC massive contact binary candidates from the OGLE survey-- bona fide sample and extended sample.}
\begin{tabular}{ |c|c|c|c|c|c|c|c|c|c| } 
    \toprule
    \multicolumn{9}{c}{\textbf{LMC Bona fide sample}}\\
    \midrule
\hline
OGLE ID & RA & Dec & Period & $m_\textrm{V,max}$ & $M_\textrm{V,max}$ & c & $f_{\Delta 
\textrm{A}}$ & Cross-references \\
\hline
OGLE-LMC-ECL-38075 & 05:47:39.45 & -68:12:18.7 & 0.933249 & 16.368 & -2.562 & 0.701 & 0.070 & -- \\
OGLE-LMC-ECL-19716 & 05:34:55.38 & -68:58:57.1 & 0.895364 & 16.345 & -2.738 & 0.704 & 0.010 & -- \\
OGLE-LMC-ECL-34618 & 05:31:22.04 & -66:33:32.5 & 1.097613 & 15.251 & -3.434 & 0.704 & 0.030 & -- \\
OGLE-LMC-ECL-19005 & 05:33:14.45 & -68:30:12.4 & 1.113130 & 14.735 & -4.141 & 0.709 & 0.069 & -- \\
OGLE-LMC-ECL-19937 & 05:35:27.02 & -69:41:48.4 & 0.999261 & 14.480 & -4.412 & 0.710 & 0.099 & -- \\
OGLE-LMC-ECL-05771 & 05:02:42.05 & -70:38:53.1 & 1.032258 & 16.160 & -2.648 & 0.711 & 0.033 & -- \\
OGLE-LMC-ECL-17758 & 05:30:33.48 & -69:05:41.0 & 0.958001 & 16.207 & -2.824 & 0.726 & 0.013 & -- \\
OGLE-LMC-ECL-37345 & 05:43:03.84 & -67:11:05.5 & 0.680809 & 15.997 & -2.713 & 0.733 & 0.013 & -- \\
OGLE-LMC-ECL-15994 & 05:27:00.82 & -68:50:09.8 & 0.676958 & 15.608 & -3.227 & 0.743 & 0.005 & -- \\
OGLE-LMC-ECL-12878 & 05:19:45.70 & -71:14:19.2 & 0.796062 & 15.713 & -3.095 & 0.746 & 0.006 & -- \\
OGLE-LMC-ECL-16532 & 05:28:06.72 & -69:01:11.7 & 0.743770 & 14.289 & -4.614 & 0.752 & 0.028 & -- \\
OGLE-LMC-ECL-05389 & 05:01:47.05 & -70:54:08.7 & 0.918771 & 16.222 & -2.553 & 0.762 & 0.025 & -- \\
OGLE-LMC-ECL-13740 & 05:21:55.17 & -67:56:11.1 & 0.871272 & 14.746 & -4.084 & 0.778 & 0.005 & -- \\
OGLE-LMC-ECL-11658 & 05:16:46.26 & -69:30:59.1 & 0.930758 & 16.099 & -2.622 & 0.785 & 0.009 & -- \\
OGLE-LMC-ECL-12982 & 05:19:59.86 & -68:15:09.2 & 0.825985 & 14.987 & -3.851 & 0.786 & 0.007 & -- \\
OGLE-LMC-ECL-06731 & 05:04:50.84 & -70:08:58.9 & 1.166866 & 15.441 & -3.334 & 0.793 & 0.035 & -- \\
OGLE-LMC-ECL-34083 & 05:29:30.71 & -65:53:22.1 & 0.695256 & 16.154 & -2.597 & 0.795 & 0.007 & -- \\
OGLE-LMC-ECL-20199 & 05:35:58.68 & -69:40:16.8 & 0.804633 & 15.937 & -2.999 & 0.800 & 0.003 & -- \\
OGLE-LMC-ECL-03838 & 04:57:30.57 & -68:49:50.4 & 0.787358 & 15.477 & -3.418 & 0.808 & 0.012 & -- \\
OGLE-LMC-ECL-18774 & 05:32:44.44 & -68:24:35.8 & 1.133080 & 14.737 & -4.558 & 0.812 & 0.008 & -- \\
OGLE-LMC-ECL-33186 & 05:25:56.53 & -67:58:12.8 & 0.636079 & 16.002 & -2.885 & 0.812 & 0.029 & -- \\
OGLE-LMC-ECL-35870 & 05:35:44.48 & -67:33:55.1 & 0.765128 & 16.090 & -3.216 & 0.830 & 0.002 & -- \\
OGLE-LMC-ECL-13761 & 05:21:57.96 & -67:57:36.5 & 0.741191 & 15.217 & -3.613 & 0.833 & 0.016 & -- \\
OGLE-LMC-ECL-14224 & 05:22:55.71 & -69:51:12.3 & 0.766442 & 15.387 & -3.445 & 0.836 & 0.011 & -- \\
OGLE-LMC-ECL-09129 & 05:10:21.64 & -67:54:17.2 & 1.039010 & 15.709 & -3.137 & 0.857 & 0.040 & -- \\
OGLE-LMC-ECL-22831 & 05:42:07.57 & -69:40:31.4 & 0.655365 & 14.992 & -3.868 & 0.863 & 0.000 & -- \\
OGLE-LMC-ECL-20081 & 05:35:45.02 & -69:39:30.9 & 0.877345 & 14.349 & -4.587 & 0.864 & 0.048 & -- \\
OGLE-LMC-ECL-36504 & 05:38:28.46 & -69:11:19.1 & 1.124145 & 14.368 & -5.253 & 0.911 & 0.026 & VFTS 352\\
\hline

    \midrule
    \multicolumn{9}{c}{\textbf{LMC Extended sample}}\\
    \midrule
\hline

OGLE-LMC-ECL-12469 & 05:18:43.39 & -68:06:53.1 & 2.579821 & 15.190 & -3.664 & 0.702 & 0.093 & -- \\
OGLE-LMC-ECL-06663 & 05:04:42.90 & -70:36:44.4 & 1.370347 & 15.650 & -3.152 & 0.703 & 0.061 & -- \\
OGLE-LMC-ECL-32685 & 05:23:58.30 & -67:36:50.5 & 2.440618 & 15.483 & -3.279 & 0.703 & 0.071 & -- \\
OGLE-LMC-ECL-02435 & 04:53:59.49 & -70:41:26.6 & 2.512764 & 15.709 & -3.232 & 0.704 & 0.012 & -- \\
OGLE-LMC-ECL-20344 & 05:36:15.92 & -69:50:12.1 & 2.852200 & 16.172 & -2.769 & 0.705 & 0.027 & -- \\
OGLE-LMC-ECL-00987 & 04:48:15.51 & -68:26:23.0 & 1.267155 & 15.358 & -3.458 & 0.707 & 0.088 & -- \\
OGLE-LMC-ECL-04397 & 04:58:59.45 & -67:29:37.3 & 2.833143 & 15.216 & -3.562 & 0.709 & 0.026 & -- \\
OGLE-LMC-ECL-12466 & 05:18:42.40 & -69:14:20.5 & 1.239323 & 15.211 & -3.523 & 0.709 & 0.071 & -- \\
OGLE-LMC-ECL-17578 & 05:30:12.12 & -69:05:26.4 & 1.884070 & 15.543 & -3.292 & 0.716 & 0.050 & -- \\
OGLE-LMC-ECL-32196 & 05:21:35.62 & -65:45:41.8 & 1.316931 & 15.213 & -3.530 & 0.716 & 0.054 & -- \\
OGLE-LMC-ECL-02646 & 04:54:33.79 & -67:01:05.9 & 1.878344 & 16.212 & -2.569 & 0.720 & 0.094 & -- \\
OGLE-LMC-ECL-07158 & 05:05:49.58 & -70:32:01.6 & 2.192314 & 14.743 & -4.035 & 0.722 & 0.086 & -- \\
OGLE-LMC-ECL-08427 & 05:08:42.98 & -68:46:09.2 & 1.334172 & 15.371 & -3.437 & 0.723 & 0.045 & -- \\
OGLE-LMC-ECL-19857 & 05:35:14.36 & -69:05:58.5 & 2.465643 & 16.245 & -3.874 & 0.724 & 0.047 & -- \\
OGLE-LMC-ECL-22111 & 05:40:23.79 & -68:29:42.4 & 1.357525 & 16.214 & -2.564 & 0.725 & 0.038 & -- \\
OGLE-LMC-ECL-19743 & 05:34:59.77 & -69:16:53.1 & 1.640856 & 15.046 & -3.950 & 0.729 & 0.010 & -- \\
OGLE-LMC-ECL-35122 & 05:33:05.82 & -67:03:19.5 & 2.239486 & 14.813 & -3.845 & 0.730 & 0.060 & -- \\
OGLE-LMC-ECL-03044 & 04:55:35.51 & -69:29:44.0 & 1.519330 & 15.589 & -3.243 & 0.743 & 0.033 & -- \\
OGLE-LMC-ECL-35479 & 05:34:22.84 & -66:53:44.3 & 2.527220 & 14.631 & -4.106 & 0.745 & 0.071 & -- \\
OGLE-LMC-ECL-06955 & 05:05:21.92 & -70:41:19.5 & 1.661544 & 15.157 & -3.547 & 0.748 & 0.056 & -- \\
OGLE-LMC-ECL-16543 & 05:28:09.04 & -69:02:43.5 & 2.411412 & 14.700 & -4.434 & 0.748 & 0.083 & -- \\
OGLE-LMC-ECL-03833 & 04:57:29.53 & -70:21:15.0 & 1.228582 & 16.291 & -2.585 & 0.748 & 0.049 & -- \\
OGLE-LMC-ECL-08338 & 05:08:32.69 & -71:13:38.5 & 1.277471 & 15.231 & -3.653 & 0.761 & 0.024 & -- \\
OGLE-LMC-ECL-02252 & 04:53:25.30 & -68:52:32.5 & 1.617946 & 14.724 & -4.236 & 0.769 & 0.056 & -- \\
OGLE-LMC-ECL-03734 & 04:57:15.67 & -68:53:44.2 & 2.195574 & 14.987 & -3.979 & 0.769 & 0.020 & -- \\
OGLE-LMC-ECL-22000 & 05:40:07.55 & -69:24:31.9 & 1.552945 & 13.187 & -5.857 & 0.794 & 0.087 & -- \\
OGLE-LMC-ECL-27516 & 04:54:56.02 & -69:00:44.8 & 1.504985 & 15.489 & -3.311 & 0.855 & 0.030 & -- \\
OGLE-LMC-ECL-02078 & 04:52:56.15 & -69:22:09.5 & 1.600595 & 15.638 & -3.314 & 0.911 & 0.071 & -- \\
\hline
\end{tabular}
\label{LMC_OGLE_CBs}
\tablefoot{Columns are: OGLE ID, Right Ascension (RA), Declination (Dec), Period (days), apparent V-band peak magnitude ($m_\textrm{V}$), absolute V-band peak magnitude ($M_\textrm{V,max}$), morph parameter (c), relative eclipse depth ($f_{\Delta \textrm{A}}$) and, cross references. The bona fide sample (28 systems) with $P<1.2$\,days aligns with our theoretical constraints for LMC CBs, while the extended sample goes up to $P=3$\,days (28 systems).}
\end{table*}

\begin{table*}
\centering
\caption{SMC massive contact binary candidates from the OGLE survey-- bona fide and extended sample.}
\begin{tabular}{ |c|c|c|c|c|c|c|c|c|c| } 
    \toprule
    \multicolumn{9}{c}{\textbf{SMC Bona fide sample}}\\
    \midrule
\hline
OGLE ID & RA & Dec & Period & $m_\textrm{V,max}$ & $M_\textrm{V,max}$ & c & $f_{\Delta \textrm{A}}$ & Cross-references \\
 \hline
OGLE-SMC-ECL-4598 & 01:02:54.38 & -72:25:00.6 & 0.685235 & 16.438 & -2.779 & 0.706 & 0.021 & -- \\
OGLE-SMC-ECL-5044 & 01:05:30.59 & -72:01:21.8 & 0.883099 & 14.880 & -4.284 & 0.729 & 0.057 & OGLE SMC-SC10 108086 \\
OGLE-SMC-ECL-4509 & 01:02:24.97 & -72:09:26.7 & 0.616745 & 15.215 & -3.895 & 0.736 & 0.050 & -- \\
OGLE-SMC-ECL-1403 & 00:48:17.96 & -73:07:19.0 & 0.868581 & 14.502 & -4.868 & 0.742 & 0.069 & -- \\
OGLE-SMC-ECL-6118 & 01:25:23.15 & -73:16:39.0 & 0.937284 & 15.138 & -4.037 & 0.742 & 0.024 & -- \\
OGLE-SMC-ECL-2774 & 00:53:34.42 & -72:42:32.8 & 0.806301 & 15.765 & -3.358 & 0.778 & 0.004 & -- \\
OGLE-SMC-ECL-5992 & 01:19:54.63 & -73:13:57.8 & 0.968954 & 16.157 & -3.082 & 0.797 & 0.034 & -- \\
OGLE-SMC-ECL-3406 & 00:56:19.27 & -72:21:04.0 & 0.663019 & 16.092 & -3.123 & 0.837 & 0.007 & SV* HV 12132 \\
\hline
    \midrule
    \multicolumn{9}{c}{\textbf{SMC Extended sample}}\\
    \midrule
\hline
OGLE-SMC-ECL-5158 & 01:06:36.70 & -72:42:39.7 & 1.315871 & 16.161 & -2.997 & 0.705 & 0.078 & -- \\
OGLE-SMC-ECL-1475 & 00:48:38.14 & -73:13:54.2 & 2.489125 & 14.601 & -4.555 & 0.706 & 0.069 & -- \\
OGLE-SMC-ECL-4024 & 00:59:34.19 & -72:46:57.9 & 1.123068 & 14.096 & -5.051 & 0.706 & 0.094 & -- \\
OGLE-SMC-ECL-3243 & 00:55:31.57 & -72:43:08.0 & 1.310807 & 15.629 & -3.487 & 0.709 & 0.037 & -- \\
OGLE-SMC-ECL-0850 & 00:45:18.20 & -73:15:23.1 & 1.002826 & 13.777 & -5.510 & 0.711 & 0.047 & -- \\
OGLE-SMC-ECL-3730 & 00:58:07.45 & -72:15:48.3 & 1.417171 & 15.085 & -4.036 & 0.716 & 0.061 & -- \\
OGLE-SMC-ECL-5272 & 01:07:31.48 & -72:19:53.0 & 1.296944 & 15.811 & -3.316 & 0.717 & 0.052 & -- \\
OGLE-SMC-ECL-6553 & 00:46:06.42 & -72:08:40.1 & 1.045789 & 16.479 & -2.642 & 0.737 & 0.043 & -- \\
OGLE-SMC-ECL-2529 & 00:52:43.03 & -72:42:24.5 & 1.074551 & 15.256 & -3.880 & 0.742 & 0.015 & -- \\
OGLE-SMC-ECL-5750 & 01:13:38.88 & -73:22:36.9 & 1.211270 & 16.201 & -3.009 & 0.745 & 0.018 & -- \\
OGLE-SMC-ECL-6016 & 01:21:01.96 & -72:20:28.2 & 2.423924 & 15.839 & -3.262 & 0.746 & 0.074 & -- \\
OGLE-SMC-ECL-5898 & 01:16:35.81 & -73:22:11.0 & 1.947295 & 16.113 & -3.104 & 0.747 & 0.071 & -- \\
OGLE-SMC-ECL-4550 & 01:02:39.95 & -72:16:32.0 & 2.175431 & 15.526 & -3.575 & 0.749 & 0.067 & -- \\
OGLE-SMC-ECL-2990 & 00:54:22.78 & -72:18:52.7 & 1.649417 & 15.701 & -3.369 & 0.760 & 0.033 & -- \\
OGLE-SMC-ECL-4690 & 01:03:21.30 & -72:05:38.2 & 2.206112 & 13.535 & -5.544 & 0.856 & 0.050 & -- \\
\hline
\end{tabular}
\label{SMC_OGLE_CBs}
\tablefoot{Columns are the same as Table~\ref{LMC_OGLE_CBs}. The bona fide sample (9 systems) with $P<1.0$\,days aligns with our theoretical constraints for SMC CBs, while the extended sample goes up to $P=3$\,days (15 systems).}
\end{table*}

\end{appendix}
\end{document}